\pdfoutput=1
\documentclass[notitlepage,prd,aps,preprintnumbers,floats,floatfix,superscriptaddress,preprintnumbers,showpacs,eqsecnum,nofootinbib]{revtex4-1}
\usepackage[utf8]{inputenc}
\usepackage{amsfonts,amsmath,amssymb,latexsym,array,afterpage,bm,float,enumitem,theorem,mathrsfs,epsfig,color,graphicx,tabularx,here,multirow,psfrag}
\newtheorem{theorem}{Theorem}
\definecolor{ultramarine}{rgb}{0.07, 0.04, 0.56}
\definecolor{cadmiumgreen}{rgb}{0.0, 0.42, 0.24}
\definecolor{indigo(dye)}{rgb}{0.0, 0.25, 0.42}
\usepackage[linktocpage=true]{hyperref}
\hypersetup{
colorlinks=true,
citecolor=ultramarine,
linkcolor=cadmiumgreen,
urlcolor=indigo(dye),
}

\newcommand{\nn}{\nonumber \\}

\newcommand{\tL}{\tilde{L}{} }
\newcommand{\f}[2]{\frac{#1}{#2}}  
\newcommand{\mk}[1]{\left( #1 \right)}

\newcommand{\be}{\begin{equation}}  
\newcommand{\ee}{\end{equation}}
\newcommand{\bem}{\begin{pmatrix}}
\newcommand{\eem}{\end{pmatrix}}

\newcommand{\pa}{\partial}

\begin{document}
\baselineskip=12pt

%<<<<<<<<<<<<< TITLE >>>>>>>>>>>>>>>%
\title{Ghost from constraints: a generalization of Ostrogradsky theorem}

%<<<<<<<<<<<<< AUTHOR >>>>>>>>>>>>>>>%
\author{Katsuki \sc{Aoki}}
\affiliation{Center for Gravitational Physics, Yukawa Institute for Theoretical Physics, Kyoto University, Kyoto 606-8502, Japan}

\author{Hayato \sc{Motohashi}$^1$\footnote{Present address: Division of Liberal Arts, Kogakuin University, 2665-1 Nakano-machi, Hachioji, Tokyo, 192-0015, Japan}}
\noaffiliation
\affiliation{Center for Gravitational Physics, Yukawa Institute for Theoretical Physics, Kyoto University, Kyoto 606-8502, Japan}

\preprint{YITP-20-06}

%<<<<<<<<<<<<< DATE >>>>>>>>>>>>>>>%
\date{\today}

%======================================%
%<<<<<<<<<<<<< ABSTRACT >>>>>>>>>>>>>>>%
%======================================%
\begin{abstract}
Ostrogradsky theorem states that Hamiltonian is unbounded when Euler-Lagrange equations are {\it higher} than second-order differential equations under the nondegeneracy assumption.
Since higher-order nondegenerate Lagrangian can be always recast into an equivalent system with at most first-order derivatives by introducing auxiliary variables and constraints, it is conceivable that the link between ghost and higher derivatives may be reinterpreted as a link between ghost and constraints and/or auxiliary variables.
We find that the latter point of view actually provides more general perspective than the former, by 
exploring the un/boundedness of the Hamiltonian for general theories containing auxiliary variables, for which Euler-Lagrange equations can be essentially second order or {\it lower} than that.
For Lagrangians including auxiliary variables nonlinearly, we derive the degeneracy condition to evade the Ostrogradsky ghost that can apply even if auxiliary variables can be solved only locally.  
For theories with constraints with Lagrange multipliers, we establish criteria for inclusion of nonholonomic (velocity-dependent) constraints leading to 
the absence of local minimum of
Hamiltonian.
Our criteria include the Ostrogradsky theorem as a special case, and can detect not only ghost associated with higher-order derivatives, but also ghost coming from lower-order derivatives in system with constraints.
We discuss how to evade such a ghost.
We also provide various specific examples to highlight application and limitation of our general arguments.
\end{abstract}

%<<<<<<<<<<<<< PACS NUMBER >>>>>>>>>>>>>>>%
%\pacs{04.60.Cf, 04.50.Gh, 04.50.-h, 11.25.-w }

% 04.40.Dg	Relativistic stars: structure, stability, and oscillations (see also 97.60.-s Late stages of stellar evolution)
% 04.50.Gh : higher-dimensional Black holes
% 04.50.-h : Higher-dimensional gravity and other theories of gravity
% 04.60.Cf : gravitational aspects of String theory
% 11.25.-w : Strings and branes
% 04.50.Kd : Modified theories of gravity
% 98.80.-k : Cosmology
% 95.36.+x : Dark energy
%\pacs{}

\maketitle

\section{Introduction}%%%%%%%%%%%%%%%%%%%%%%%%%%%%%%%%%%%%%%%%%
\label{sec_intro}

Constraints are ubiquitous in physics.  This is because in general Lagrangians involve not only dynamical variables but also nondynamical ones, whose derivatives do not appear in the Lagrangian.  The Euler-Lagrange equations for such variables yield constraint equations, which contain nondynamical and dynamical variables, and/or their derivatives.  If they can be solved for nondynamical variables, one can eliminate them by substituting their solutions back into the Lagrangian.  If they restrict evolution of dynamical variables, they reduce the number of degrees of freedom in phase space.
One might think that constraints are not essential since they can be eliminated by substituting solutions of the constraints to Lagrangian, with the help of a gauge fixing condition if a constraint is first class, and then an equivalent system without constraints may be obtained in principle. 
Nevertheless, in many systems constraints cannot be solved explicitly. 
Furthermore, to manifest a symmetry of a system, nondynamical variables and constraints, say the lapse and the Hamiltonian constraint in theories of gravity, play important roles. 
Also, when a theory contains some dynamical degrees of freedom that are sufficiently massive and hence can be approximated as being nondynamical, the theory can be regarded as an effective field theory (EFT) with nondynamical degrees freedom below some cutoff energy scale.
In such a case, the EFT action with the nondynamical ones would provide more hints of the original theory than the one after integrating them out.
Therefore, it would be advantageous to retain constraints in a defining action of a theory, and any universal understanding of constraints must be helpful for a deep understanding of physics.

The present study is particularly motivated by the recent progress of theories of modified gravity.
Counting number of degrees of freedom is crucial in modified gravity, where extra degrees of freedom are introduced to General Relativity so that they are responsible for primordial inflation or late-time accelerated expansion of the Universe and/or serve as a framework for testing gravity in the strong field regime such as the vicinity of black holes~\cite{Clifton:2011jh,Joyce:2014kja,Berti:2015itd,Koyama:2015vza}.  To construct sensible theories, one needs to tame the additional degrees of freedom so that they do not cause instabilities such as ghost instability.  
In fact, the constraints play two important roles to eliminate the Ostrogradsky ghosts associated with higher-derivative theories: rewriting a higher-derivative theory to an equivalent lower-derivative theory allows us a systematic study, and imposing appropriate constraints remove ghost degrees of freedom, allowing a construction of ghost-free theory.  
Since the second role is more directly related to the elimination the ghost degrees of freedom, one may think that the first role is not essential. 
However, the first role actually tells us a close interplay between ghosts and constraints.

To revisit the first role, let us begin with the Ostrogradsky's theorem~\cite{Ostrogradsky:1850fid,Woodard:2015zca}.
The Ostrogradsky's theorem states as follows:  
\begin{theorem}[Ostrogradsky theorem] \label{Ostro_th}
Let a Lagrangian involves $n$-th order finite time derivatives of variables. If $n\geq 2$ and the Lagrangian is nondegenerate with respect to the highest-order derivatives, 
the Hamiltonian of this system linearly depends on a canonical momentum.
\end{theorem}
The linear dependency of the canonical momentum implies the following local and global pathological properties: the Hamiltonian has no local minimum and the Hamiltonian is unbounded from below. 
Furthermore, it was shown recently that the unbounded Hamiltonian outlasts even after quantization~\cite{Raidal:2016wop,Smilga:2017arl,Motohashi:2020psc}.
If the assumption of the Ostrogradsky theorem is satisfied, the Euler-Lagrange equations form a system of $2n$-th order differential equations.
While the Ostrogradsky theorem focuses only on the ghost degrees of freedom associated with the highest-order derivatives in the Euler-Lagrange equations, in general non-highest but higher-order derivatives also lead to an unbounded Hamiltonian~\cite{Motohashi:2014opa}.
Therefore, in general, higher-derivative Lagrangian having higher-order system of Euler-Lagrange equations 
suffers from unbounded Hamiltonian.
This is considered to be the reason why the laws of physics are described by second-order differential equations rather than higher-order ones. 
However, one can always rewrite a higher-derivative Lagrangian $L=L(\phi,\dot{\phi},\ddot{\phi},\cdots)$ as an equivalent Lagrangian $L_{\rm eq}=L(q^I,\dot{q}^I)+\lambda^a C_a(q^I,\dot{q}^I)$ up to first-order time derivatives by introducing Lagrange multipliers $\lambda^a$ and constraints $C_a=q^a-\dot q^{a-1}$.
In this case, un/boundedness of Hamiltonian is determined by the form of $L(q^I,\dot{q}^I)$ and $C_a$.
The higher-derivative nature of the original Lagrangian is encoded into $C_a$ in the equivalent lower-order Lagrangian. 
Then, a natural question arises: Does the Ostrogradsky instability exist only for this particular $C_a$? 
If this is not the case, a Lagrangian with auxiliary variables and constraints that cannot be recast to higher-order system could also suffer from the unbounded Hamiltonian, for which the Euler-Lagrange equations form a second- or lower-order system of differential equations.
Then, the ghost would be associated with auxiliary variables and/or constraints rather than higher-order derivatives. 
Therefore, considering Lagrangian with auxiliary variables and constraints would provide a more general perspective than the Ostrogradsky theorem.

On the other hand, as mentioned above, the second role has been focused as the essence of the elimination of the Ostrogradsky ghosts.  
Chen et al.\ \cite{Chen:2012au} investigated nondegenerate higher-derivative Lagrangian with constraints, and concluded that the Ostrogradsky instability can only be removed by the addition of constraints if the original theory's phase space is reduced.
Together with the first role of the constraints, Ostrogradsky ghosts in arbitrary higher-derivative theories can be systematically eliminated by imposing an appropriate set of constraints~\cite{Langlois:2015cwa,Motohashi:2016ftl,Motohashi:2017eya,Motohashi:2018pxg}.
This procedure was applied to a construction of degenerate higher-order scalar-tensor (DHOST) theories~\cite{Langlois:2015cwa,BenAchour:2016fzp}, where the degeneracy conditions are imposed to guarantee the existence of the constraints to eliminate the Ostrogradsky ghosts. 
These results are reasonable since ghost degrees of freedom in phase space are removed by a certain set of constraints.
However, from more general point of view, it is not clear whether adding constraints is always a good thing.
In principle, constraints would eliminate healthy degrees of freedom rather than ghosts, or would not reduce phase space dimension.
It is not clear what happens for these cases.
One interesting example is a class of scalar-tensor theories generated by non-invertible metric transformation, known as mimetic gravity~\cite{Chamseddine:2013kea} (see also \cite{Lim:2010yk,Gao:2010gj,Capozziello:2010uv}),
which has a rich phenomenology for cosmology and astrophysics 
(for a recent review, see \cite{Sebastiani:2016ras}).  
The instability issue of this class of theories has been studied extensively~\cite{Takahashi:2017pje,Langlois:2018jdg,Ramazanov:2016xhp,Ganz:2018mqi}.
It is also known that the action of this class can be rewritten in the form with non-holonomic constraint.
Nevertheless, the interplay between the constraint and the instability has not been clarified yet.

To address these questions, in this paper we consider general Lagrangian with auxiliary variables and constraints
We find that the Ostrogradsky instability, which has been regarded to originate from the higher-derivative nature of the Lagrangian, is more generally related to constraints that does not reduce the phase space dimension.
Indeed, we find various examples that exhibit the same type of unbounded Hamiltonian due to terms linear in canonical variables. 
The main theorem we prove in the present paper is as follows: 
\begin{theorem}\label{main_th}
Let a Lagrangian $L=L_0(\dot{q}^I,q^I)+\lambda^a C_a(\dot{q}^I,q^I)$ be equipped with constraints $C_a=0$ via the Lagrange multipliers $\lambda^a$, where the dimensions of $q^I$ and $\lambda^a$ are $n$ and $m$ ($n\geq m$), respectively. 
The Hamiltonian has neither local minimum nor maximum
if the following three conditions are satisfied: 
\ref{con1}.~The Lagrangian is nondegenerate with respect to $\dot{q}^I$. 
\ref{con2}.~The constraints do not reduce the phase space dimension of $q^I$. 
\ref{con3}.~Either a)~the system does not allow a solution $\dot{q}^I=0=\f{\pa L}{\pa q^I}$ under $C_a=0$, or b)~at least one of the constraints has $q^I$ dependency at the solutions $\dot{q}^I=0=\f{\pa L}{\pa q^I}$ under $C_a=0$.
\end{theorem}
The condition \ref{con3} consists of two cases, \ref{con3}-a and \ref{con3}-b. The condition \ref{con3}-a deals with the case that a particle cannot stop due to a constraint and then the Hamiltonian does not admit a stationary point. 
Although the Hamiltonian does not have local minimum nor maximum, this case is not pathological on the same footing with the Ostrogradsky ghosts.
On the other hand, the Ostrogradsky-like system is found when the condition \ref{con3}-b holds. As we will show in \S\ref{sec_aux22}, the condition \ref{con3}-b accompanying the conditions \ref{con1} and \ref{con2} conclude that all of stationary points of the Hamiltonian are saddle points, i.e.\ there is a mode that yields a lower energy state than any stationary point. Therefore, the system does not admit a (meta)stable state and there exists a ghostly degree of freedom. Since the systems discussed in Theorem \ref{main_th} include not only systems that are equivalent to higher derivative systems but also others, it is reasonable to view Theorem \ref{main_th} as a generalization of the Ostrogradsky theorem.\footnote{Note that the consequences of Theorem \ref{Ostro_th} and Theorem \ref{main_th} are slightly different. Strictly speaking, the Ostrogradsky theorem concludes that the Hamiltonian is linear in a canonical momentum and then not bounded from above and below while the Theorem~\ref{main_th} still allows a bounded Hamiltonian whose infimum is located on the boundary in phase space. 
We will discuss this point in \S\ref{sec_canonical}.
} 
The violation of, at least, one of the conditions~\ref{con1}--\ref{con3} is necessary (but not sufficient) condition to find a local minimum of the Hamiltonian.
We will show various examples with a pathological Hamiltonian attributed to adding constraints and those with a healthy Hamiltonian due to violation of the conditions.

% ==================== Table ====================
\begin{table}[t]
\centering
\begin{tabular}{|l|l|l|l|} \hline 
 Theory & General argument & Example with ghost  & Example without ghost \\ \hline 
 $L(\ddot{\phi}_0,\dot{\phi}_0,\phi_0,\dot{q},q)$ & \S\ref{sec_deg} (see also \S\ref{sec_ex2})&  &
 \\ 
  $L(\ddot{\phi}_0,\dot{\phi}_0,\phi_0,\dot{q},q,\xi^i )$ &  \S\ref{sec_aux11} & & \S\ref{sec_aux12} 
  \\ 
 $L_0(\dot{q}^I,q^I)+\lambda^a C_a(q^I)$ & \S\ref{sec_aux21} &  &
 \\  
 $L_0(\dot{q}^I,q^I)+\lambda^a C_a(\dot{q}^I,q^I)$ & \S\ref{sec_aux22} & \S\ref{sec_ex1} \eqref{H_uns}, \S\ref{sec_ex2}, \S\ref{sec_ex4} \eqref{ex_vio3_ghost} & \S\ref{sec_ex4} \eqref{ex_vio3_GF}
 \\
 $L_0(\dot{q}^I,q^I,\xi^i)+\lambda^a C_a(\dot{q}^I,q^I,\xi^i)$ & \S\ref{sec_aux23}, \S\ref{sec_aux24} & & \S\ref{sec_ex1} \eqref{H_sta}, \S\ref{sec_ex3}
 \\ \hline 
\end{tabular}
\caption{Theories we address in the present paper, where $\phi_0$, $q^I$ are dynamical variables, $\xi^i$ are solvable auxiliary variables, and $\lambda^a$ are Lagrange multipliers for constraints $C_a$. See also the notation summarized in \S\ref{sec_over2}. }
\label{tab_1}
\end{table}
% ==================== Table ====================

The above consideration suggests a rich structure of systems with nondynamical variables and/or constraints.
The purpose of the present paper is to establish a deeper understanding of these systems, which shall be highlighted in \S\ref{sec_over1} by a suggestive example.   
In particular, we prove Theorem~\ref{main_th} as a generalization of Theorem~\ref{Ostro_th} known as the Ostrogradsky theorem.
To this end, we shall classify theories with nondynamical variables and/or constraints, for each of which we address general arguments as well as specific examples.
The classification and the structure of the rest of the paper is summarized in Table~\ref{tab_1}.
In \S\ref{sec_deg}, we review the derivation of the degeneracy condition to eliminate the Ostrogradsky ghosts from general Lagrangian involving higher-order derivatives without auxiliary variables.
In \S\ref{sec_aux1} and \S\ref{sec_aux2}, we investigate Lagrangian with auxiliary variables and present specific examples.
First, in \S\ref{sec_aux1}, we consider the case where constraint equations for auxiliary variables are solvable for the auxiliary variables, and derive degeneracy condition to evade the Ostrogradsky ghosts.
Second, in \S\ref{sec_aux2}, we address the case where auxiliary variables are Lagrange multipliers and prove Theorem~\ref{main_th}. 
We shall also consider a way out from the Theorem~\ref{main_th} in \S\ref{sec_aux23} by introducing solvable auxiliary variable. 
In \S\ref{sec_ex} various concrete examples with Lagrange multipliers are studied.
\S\ref{sec_conc} is devoted to conclusion.

\section{Overview and notation}%%%%%%%%%%%%%%%%%%%%%%%%%%%%%%%%%%%%%%%%%
\label{sec_over}

\begin{figure}[t]
\centering
\includegraphics[width=.8\textwidth]{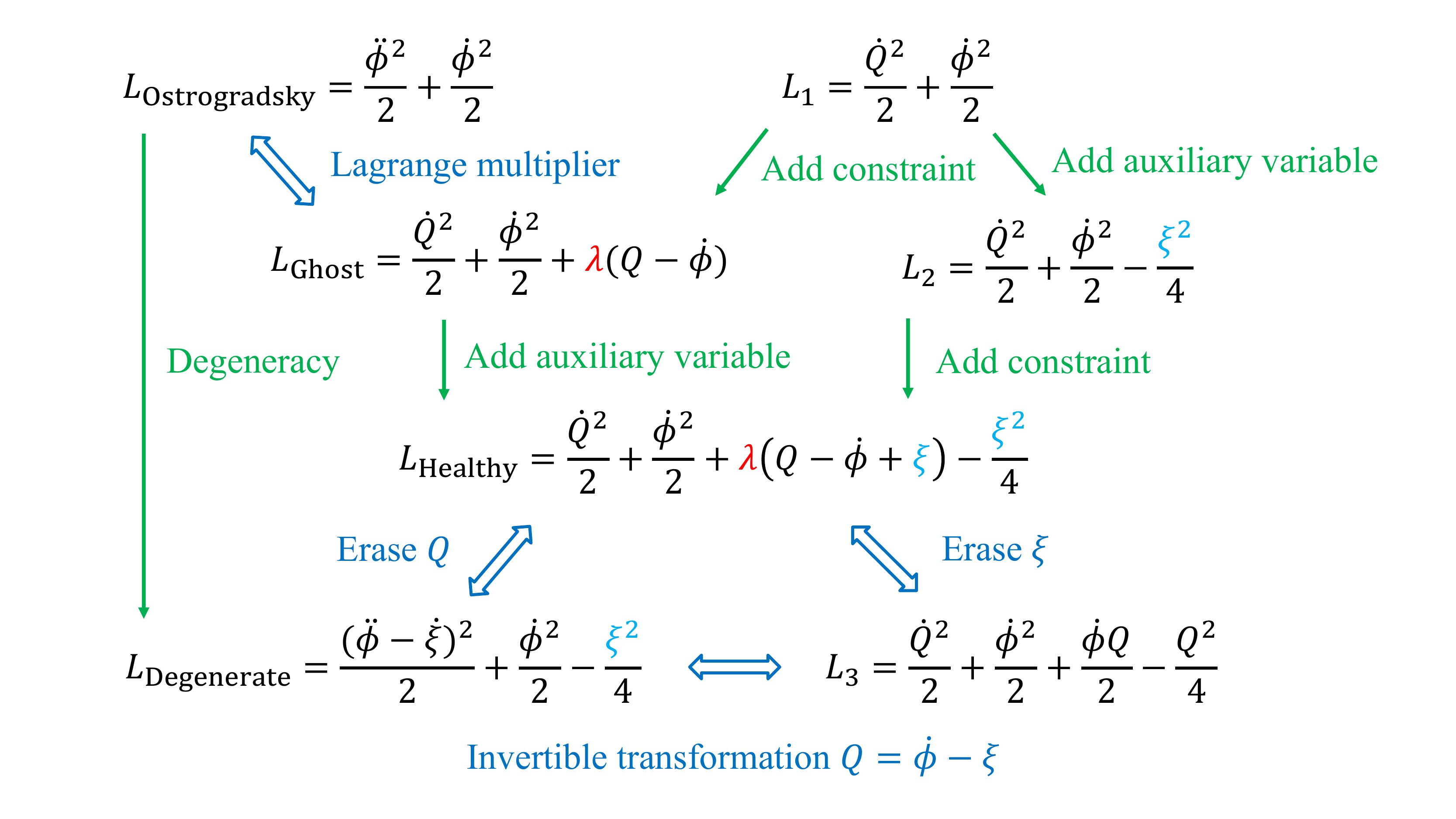}
\caption{Interplay between various Lagrangians with or without ghost linked by virtue of nondynamical variables $\xi$ and $\lambda$ as well as constraints, which provides us several different perspectives.
$L_{\text{Ostrogradsky}}$ and $L_{\text{Ghost}}$ suffer from an unbounded Hamiltonian due to a term linear in canonical momentum, which can be cured either promoting it to degenerate Lagrangian or introducing solvable auxiliary variable. 
$L_{\text{Degenerate}}$, $L_{\text{Healthy}}$, as well as $L_3$ are equivalent healthy theories via an invertible transformation or plugging a solution of the constraint.
For $L_1\to L_{\text{Ghost}}$, adding a constraint introduces a ghost degree of freedom, suggesting that adding constraint is not always a good thing.  See \S\ref{sec_over1} and \S\ref{sec_ex3} for more details.
}
\label{fig:Lags}
\end{figure}

In \S\ref{sec_over1}, as an overview of the present paper, we provide a pedagogical example on the interplay between various Lagrangians that are linked by virtue of nondynamical variables and constraints (see Fig.~\ref{fig:Lags} for a summary).
We shall revisit this example in \S\ref{sec_ex3}.
In \S\ref{sec_over2}, we summarize our notation which we use throughout the present paper.

\subsection{Interplay between Lagrangians}%%%%%%%%%%%%%%%%%%%%%
\label{sec_over1}

% ==================== Figure ====================
% ==================== Figure ====================

Let us consider a system of two free particles:
\begin{align} \label{ex_intro1}
    L_1=\frac{1}{2}\dot{Q}^2+\frac{1}{2}\dot{\phi}^2
    ,
\end{align}
which is clearly free from any instability. However, when a nonholonomic (velocity-dependent) constraint $Q-\dot{\phi}=0$ is added to the system, the Lagrangian,
\begin{align} \label{ex_intro2}
    L_{\rm Ghost}=\frac{1}{2}\dot{Q}^2+\frac{1}{2}\dot{\phi}^2+\lambda (Q-\dot{\phi})
    ,
\end{align}
is equivalent to a higher-derivative Lagrangian 
\begin{align} \label{ex_intro3}
    L_{\rm Ostrogradsky}=\frac{1}{2}\ddot{\phi}^2+\frac{1}{2}\dot{\phi}^2
\end{align}
since the constraint can be solved as $Q=\dot{\phi}$. 
From Theorem~\ref{Ostro_th}, the Lagrangian~\eqref{ex_intro3} suffers from the Ostrogradsky ghost.
Hence, the ``constrained'' Lagrangian~\eqref{ex_intro2} also suffers from the Ostrogradsky ghost. Note that the phase space dimension of both \eqref{ex_intro1} and \eqref{ex_intro2} is four; the nonholonomic constraint $Q-\dot{\phi}=0$ does not eliminate the degree of freedom of either $Q$ or $\phi$. Furthermore, the constraint $Q-\dot{\phi}=0$ has the $Q$ dependency. Therefore, Theorem~\ref{main_th} also concludes the system \eqref{ex_intro2} exhibits a pathological Hamiltonian.

Although for this theory both of Theorem~\ref{Ostro_th} and Theorem~\ref{main_th} works for \eqref{ex_intro3} and \eqref{ex_intro2}, respectively, Theorem~\ref{main_th} can apply to a wider class of theories than Theorem~\ref{Ostro_th}.
Actually, a pathological Hamiltonian can show up even if a Lagrangian cannot be rewritten as a higher-derivative theory.
For such a case, we can still exploit Theorem~\ref{main_th} while Theorem~\ref{Ostro_th} does not apply.
We shall provide specific examples for such a case in \S\ref{sec_aux22}, \S\ref{sec_aux24} and \S\ref{sec_ex1}.
Hence, Theorem~\ref{main_th} provides us a more general and robust point of view to unveil the origin of ghost as nonholonomic constraints that do not reduce the phase space dimension, rather than Theorem~\ref{Ostro_th} interpreting the origin of ghost as the nondegeneracy of the Lagrangian with respect to the highest-order derivatives.

We then consider 
\begin{align} \label{ex_intro4}
    L_2=\frac{1}{2}\dot{Q}^2+\frac{1}{2}\dot{\phi}^2-\frac{1}{4}\xi^2
    ,
\end{align}
which has three variables but only has two dynamical degrees of freedom since $\xi$ is nondynamical. 
The Lagrangian~\eqref{ex_intro4} would be one of the simplest degenerate Lagrangian in the sense that $\xi$ does not have a kinetic term.  
Similarly to the previous example, we add a nonholonomic constraint and consider a new Lagrangian
\begin{align} \label{ex_intro5}
     L_{\rm Healthy}=\frac{1}{2}\dot{Q}^2+\frac{1}{2}\dot{\phi}^2-\frac{1}{4}\xi^2 + \lambda(Q-\dot{\phi}+\xi)
    ,
\end{align}
which can be also obtained by introducing nondynamical variable $\xi$ to the Lagrangian~\eqref{ex_intro2}, which suffers from the ghost.
Contrary to \eqref{ex_intro2}, the Lagrangian \eqref{ex_intro5} is free from the Ostrogradsky ghost: the ghost is exorcised by inclusion of the nondynamical variable $\xi$, namely the violation of the condition \ref{con1}. Indeed, the constraint can be solved as $Q=\dot{\phi}-\xi$ and then Lagrangian becomes a degenerate higher-order theory
\begin{align} \label{ex_intro6}
    L_{\rm Degenerate}=\frac{1}{2}(\ddot{\phi}-\dot{\xi})^2+\frac{1}{2}\dot{\phi}^2-\frac{1}{4}\xi^2.
\end{align}
A more straightforward way to see the ghost-freeness is solving the constraint as $\xi=\dot{\phi}-Q$. Substituting it, we obtain the Lagrangian
\begin{align} \label{ex_intro7}
    L_3=\frac{1}{4}\dot{Q}^2+\frac{1}{2}\dot{\phi}^2+\frac{1}{2}\dot{\phi} Q -\frac{1}{4}Q^2,
\end{align}
which is clearly free from the Ostrogradsky instability. Two Lagrangians \eqref{ex_intro6} and \eqref{ex_intro7} are related via $Q=\dot{\phi}-\xi$, which can be regarded as the invertible transformation.

Although one can easily see the relations between three apparently different Lagrangians, $L_{\rm Healthy}$, $L_{\rm Degenerate}$, and $L_3$ in this simple example, it is not straightforward to either solve constraints explicitly or take an invertible transformation, in general. Even if the system is equivalent, attention has to be paid to the choice of variables. In the context of modified gravity, the choice of variables may have a certain physical meaning such as the Einstein frame metric and the Jordan frame metric where the former one is the metric of which kinetic term is the Einstein-Hilbert action whereas the latter one is the metric which the matter fields couple to. Furthermore, the change of variables has been used to find new theories of modified gravity~\cite{Zumalacarregui:2013pma,Aoki:2018zcv}.
Of course, theories related through a change of variables, or more generally an invertible transformation, share the same dynamics~\cite{Takahashi:2017zgr}.
However, one can regard them as different theories by a choice of theory or frame that matter couples to; for instance, although two Lagrangians $L_{\rm Degenerate}$ and $L_3$ are equivalent, two theories become different if one puts another variable depending on whether it couples to $\xi$ or $Q$.
The previous studies focused on the relation between $L_{\rm Degenerate}$ and $L_3$.
In the present paper, on the other hand, we shall focus on the relation between $L_{\rm Healthy}$ and others and discuss general properties of Lagrangian with auxiliary variables.

\subsection{Notation}%%%%%%%%%%%%%%%%%%%%%
\label{sec_over2}

Throughout the present paper, we use the following notation. Our Lagrangian $L$ depends on three kinds of time-dependent variables, $q^I, \lambda^a, \xi^i$.  The variables $q^I$ denote dynamical variables, whose first-order derivatives are included in the Lagrangian, whereas $\lambda^a$ and $\xi^i$ denote nondynamical variables, whose derivatives do not appear in the Lagrangian and dynamics are determined by $q^I$. The variables $\lambda^a$ are Lagrange multipliers, appearing in the Lagrangian only linearly.  On the other hand, the variables $\xi^i$ are solvable nondynamical variables by the use of their equations of motion, i.e.\ the Lagrangian is nonlinear in $\xi^i$.  Unless otherwise specified, the indices run over as follows:
\begin{align}
    I,J,K,\cdots &\in (1,2,\cdots, n), \\
    a,b,c,\cdots &\in (1,2,\cdots, m), \\
    i,j,k,\cdots &\in (1,2,\cdots, \ell),
\end{align}
with $m\leq n+\ell$.
Therefore, the general Lagrangian discussed in this paper is given by
\begin{align}
    L=L_0(\dot{q}^I,q^I,\xi^i)+\lambda^a C_a(\dot{q}^I,q^I,\xi^i). \label{L_general}
\end{align}
We shall investigate subclasses of this model in order (see Table~\ref{tab_1}).
We also use the following notations to express the derivatives of the Lagrangian with respect to the variables:
\begin{align} \label{deriv1}
    L_I=\frac{\partial L}{\partial \dot{q}^I},\quad 
    L_i=\frac{\partial L}{\partial \xi^i}, \quad
    L_a=\frac{\partial L}{\partial \lambda^a},\quad
    L_{IJ}=\frac{\partial^2 L}{\partial \dot{q}^I \partial \dot{q}^J},
\end{align}
and so on. Since the Lagrangian is supposed to be nonlinear in $\xi^i$, we assume
\begin{align}
    {\rm det}(L_{ij}) \neq 0.
\end{align}
Also, $L^{IJ}$ and $L^{ij}$ denote the inverse matrices of $L_{IJ}$ and $L_{ij}$, respectively.
When either $n,m$ or $\ell$ (i.e.~the number of $q^I,\lambda^a,\xi^i$) is 1, we simply omit the index of the corresponding variable. The derivatives of the Lagrangian are then denoted by, e.g.
\begin{align} \label{deriv2}
L_{\dot{q}}=\frac{\partial L}{\partial \dot{q}},\quad L_{\xi}=\frac{\partial L}{\partial \xi}
.
\end{align}

The conjugate pairs are denoted by
\begin{align}\label{conj1}
    (q^I,p_I), \quad (\xi^i,\varpi_i), \quad (\lambda^a,\pi_a).
\end{align}
Namely, $p_I\equiv \partial L/\partial \dot q^I$ and so on.
By definition, we always have the primary constraints
\begin{align}\label{pri-gen}
    \varpi_i \approx 0, \quad \pi_a \approx 0.
\end{align}
Throughout the present paper, we assume the Lagrangian is not degenerate in terms of $q^I$, i.e.\
\begin{align}
    {\rm det}(L_{IJ}) \neq 0.
\end{align}
Hence, there are no additional primary constraints and the total Hamiltonian of \eqref{L_general} is given by
\begin{align}
    H_{\rm tot}=\dot{q}^I p_I - L +\zeta^a \pi_a+\zeta^i \varpi_i
\end{align}
where $\zeta^a$ and $\zeta^i$ are the Lagrange multipliers to implement the primary constraints~\eqref{pri-gen}.
In the Hamiltonian, $\dot{q}^I$ are understood as the solutions of the equations
\begin{align}
    p_I = L_I(\dot{q}^J,q^J,\lambda^a,\xi^i).
\end{align}
Hence, $\dot{q}^I$ are generally functions of $q^J,p_J,\lambda^a$ and $\xi^i$.

While the Lagrangian~\eqref{L_general} contains at most first-order derivatives, it implicitly includes higher-derivative theories by virtue of the Lagrange multipliers. To make this point clearer, we shall use $\phi_0$ to explicitly denote a higher-derivative variable.
Let us consider a higher-derivative Lagrangian with at most $N+1$-th order derivatives
\begin{align}
    L=L(\phi_0^{(N+1)},\phi_0^{(N)},\cdots,\dot{\phi}_0,\phi_0).
\end{align}
This Lagrangian is equivalent to
\begin{align}
    L_{{\rm eq}1}=L(\dot{\phi}_a,\dot{\phi}_0,\phi_0)+\sum_{a=1}^{N}\lambda^a (\phi_a-\dot{\phi}_{a-1})
    ,\label{phi_eq1}
\end{align}
or
\begin{align}
    L_{{\rm eq}2}&=L(\dot{\phi}_{2a}, \phi_{2a},\dot{\phi}_0,\phi_0)+\sum_{a=1}^{[(N+1)/2]} \lambda^{2a} (\phi_{2a}-\ddot{\phi}_{2(a-1)})\notag\\
    &=L(\dot{\phi}_{2a}, \phi_{2a},\dot{\phi}_0,\phi_0)+\sum_{a=1}^{[(N+1)/2]} ( \lambda^{2a} \phi_{2a}+\dot{\lambda}^{2a} \dot{\phi}_{2(a-1)} ),
    \label{phi_eq2}
\end{align}
by the use of the auxiliary variables $\phi_a$ and the Lagrange multipliers $\lambda^a$.
The subscripts represent the order of the derivative when the constraints are solved: $\phi_0$ is the original variable itself whereas $\phi_1=\dot{\phi}_0, \phi_2=\ddot{\phi}_0$ and so on. 
While both of $L_{{\rm eq}1}$ and $L_{{\rm eq}2}$ describe the same theory and belong to the general form~\eqref{L_general},
they fall into different classes, since $L_{{\rm eq}1}$ contains a multiplier but $L_{{\rm eq}2}$ does not.
Note that when the higher derivative Lagrangian does not contain an odd number of derivatives, the second form \eqref{phi_eq2} contains an auxiliary variable; for instance, the equivalent form to the Lagrangian with at most second order derivative is
\begin{align}
    L_{{\rm eq}2}&=L(\phi_2,\dot{\phi}_0,\phi_0)+\lambda(\phi_2-\ddot{\phi}_0)\notag\\
    &=L(\phi_2,\dot{\phi}_0,\phi_0)+\lambda \phi_2+ \dot\lambda \dot{\phi}_0,
\end{align}
in which $\phi_2$ is an auxiliary variable.
We shall deal with various types of Lagrangians, which are summarized in Table~\ref{tab_1}.
$L_{{\rm eq}1}$ and $L_{{\rm eq}2}$ are classified into the fourth and second class in Table~\ref{tab_1}, respectively.
The equivalent action of the form $L_{{\rm eq}1}$ is particularly useful to discuss a generic Lagrangian with at most $N+1$-th order derivatives. On the other hand, the second equivalent action \eqref{phi_eq2} is useful to see the existence of the ghost in the Lagrangian formalism since the kinetic matrix of \eqref{phi_eq2} is clearly not positive definite, but the second form will be used only for the Lagrangian with at most second order derivatives in \S\ref{sec_deg}.

\section{Degenerate theory without auxiliary variable}%%%%%%%%%%%%%%%%%%%%%%%%%%%%%%%%%%%%%%%%%
\label{sec_deg}

In this section, following \cite{Motohashi:2016ftl}, we briefly review how to eliminate the Ostrogradsky ghosts in general higher-order theory without auxiliary variable by imposing degeneracy condition.
We consider a general higher-order Lagrangian 
\be L=L(\ddot{\phi}_0,\dot{\phi}_0,\phi_0,\dot{q},q), \label{L1} \ee 
for $\phi_0=\phi_0(t)$ and $q=q(t)$. 
Note that in principle all the variables can be dynamical, and none of them are a priori auxiliary variables.
In principle this system has three degrees of freedom, one of which is associated with an Ostrogradsky ghost.
By imposing a certain condition, known as the degeneracy condition, we can eliminate the unwanted ghost degree of freedom.

We shall use an equivalent form \eqref{phi_eq2} by the use of the additional variables $\phi_2$ and $\lambda$. The (non)existence of the Ostrogradsky ghost in the other equivalent form \eqref{phi_eq1} will be discussed in \S\ref{sec_ex2}.  
For derivatives of the Lagrangian, we use the similar notation as \eqref{deriv2}.
For the following, we assume that $L_{\phi_2\phi_2}\ne 0$ and $L_{\dot{q}\dot{q}}\neq 0$.
If the first assumption $L_{\phi_2\phi_2}\ne 0$ is not satisfied, by integration by parts, the Lagrangian is equivalent to the one involving at most first-order derivatives.
Needless to say, since \eqref{phi_eq2} is obtained by replacing $\ddot{\phi}_0$ with $\phi_2$, the condition $L_{\phi_2\phi_2}\neq 0$ means $\frac{\partial^2}{\partial \ddot{\phi}_0^2}L(\ddot{\phi}_0,\dot{\phi}_0,\phi_0,\dot{q},q) \neq 0$.
The second condition $L_{\dot{q}\dot{q}}\neq 0$ guarantees that $q$ is a dynamical degree of freedom.

While in \cite{Motohashi:2016ftl} (and in \eqref{phi_eq1}) we replaced $\dot\phi_0$ by another variable, in \eqref{phi_eq2} we replaced $\ddot\phi_0$ instead.
In \eqref{phi_eq1}, the variable $\lambda$ is a nondynamical variable but $\phi_1$ is dynamical. On the other hand, in \eqref{phi_eq2} the variable $\lambda$ is dynamical while the variable $\phi_2$ is a nondynamical variable.
The Euler-Lagrange equation for $\phi_2$ is given by 
\begin{align}
L_{\phi_2}+\lambda =0. \label{Qeq}
\end{align}
From the implicit function theorem, under the assumption $L_{\phi_2\phi_2}\ne 0$, we can solve~\eqref{Qeq} as $\phi_2=\phi_{\rm sol}(\dot{\phi}_0,\phi_0,\dot{q},q,\lambda)$.
Below we erase $\phi_2$ and consider the Lagrangian
\be \label{Leq} L_{\rm er} = L(\phi_{\rm sol}(\dot{\phi}_0,\phi_0,\dot{q},q,\lambda),\dot{\phi}_0,\phi_0,\dot{q},q)+\lambda \phi_{\rm sol}(\dot{\phi}_0,\phi_0,\dot{q},q,\lambda)+ \dot{\lambda} \dot{\phi}_0. \ee
Substituting $\phi_2=\phi_{\rm sol}$ into \eqref{Qeq} and taking derivatives, we obtain the following formulae
\begin{align}
\frac{\partial \phi_{\rm sol}}{\partial q}&=-\left. \frac{L_{\phi_2 q}}{L_{\phi_2 \phi_2}}\right|_{\phi_2=\phi_{\rm sol}} , \quad
\frac{\partial \phi_{\rm sol}}{\partial \phi_0}=-\left. \frac{L_{\phi_2 \phi_0}}{L_{\phi_2 \phi_2}}\right|_{\phi_2=\phi_{\rm sol}} ,
\label{f1}
\\
\frac{\partial \phi_{\rm sol}}{\partial \dot{q}}&=-\left. \frac{L_{\phi_2 \dot{q}}}{L_{\phi_2 \phi_2}}\right|_{\phi_2=\phi_{\rm sol}} , \quad
\frac{\partial \phi_{\rm sol}}{\partial \dot{\phi}_0}=-\left. \frac{L_{\phi_2 \dot{\phi}_0}}{L_{\phi_2 \phi_2}}\right|_{\phi_2=\phi_{\rm sol}} , \label{f2}
\\
\frac{\partial \phi_2}{\partial \lambda}&=-\left. \frac{1}{L_{\phi_2 \phi_2}} \right|_{\phi_2=\phi_{\rm sol}}. \label{f3}
\end{align}

From the Lagrangian~\eqref{Leq}, the canonical momenta are given by 
\begin{align}
\pi&=\dot{\phi}_0, \\
p_{\phi}&=\dot{\lambda}+L_{\dot{\phi}_0}|_{\phi_2=\phi_{\rm sol}}, \\
p&=L_{\dot{q}}|_{\phi_2=\phi_{\rm sol}}, \label{primary_c}
\end{align}
where $p_{\phi}$ is the conjugate of $\phi_0$ and we used \eqref{Qeq}.
Clearly, $\dot{\lambda}$ and $\dot{\phi}_0$ can be always solved in terms of canonical variables. 
If $\dot{q}$ is also solvable, the corresponding Hamiltonian is
\begin{align} \label{Hunb}
H= \pi p_{\phi}+\dot{q}p-\lambda \phi_{\rm sol}(\dot{\phi}_0,\phi_0,\dot{q},q,\lambda)-L|_{\phi_2=\phi_{\rm sol}(\dot{\phi}_0,\phi_0,\dot{q},q,\lambda)}.
\end{align}
where it is understood that $\dot{\phi}_0=\pi$ and $\dot{q}=\dot{q}(p,\pi,\phi_0,q,\lambda)$.
In particular, $\dot{\phi}_0$ and $\dot{q}$ are independent of $p_{\phi}$. 
Hence, $p_{\phi}$ appears only in the first term and the Hamiltonian is linear in $p_{\phi}$. 
As a result, the Hamiltonian is unbounded from below, which manifests the existence of the Ostrogradsky ghost.

To obtain a ghost-free theory, one needs to assume that $\dot{q}$ is not solvable, i.e.\
\be \frac{\partial}{\partial \dot{q}} (L_{\dot{q}}|_{\phi_2=\phi_{\rm sol}}) = 0. \ee
This requirement is equivalent to impose
\begin{align} \label{degcon}
\left[ L_{\dot{q}\dot{q}}L_{\phi_2\phi_2}-L_{\phi_2 \dot{q}}^2 \right]_{\phi_2=\phi_{\rm sol}}= 0.
\end{align}
From the assumptions $L_{\dot{q}\dot{q}}\ne 0$ and $L_{\phi_2\phi_2}\ne 0$, $L_{\phi_2\dot{q}}\ne 0$ is necessary to satisfy the condition~\eqref{degcon}.

The condition~\eqref{degcon} is nothing but the degeneracy condition~\cite{Motohashi:2016ftl},
\begin{align}\label{st-deg}
L_{\ddot{\phi}_0\ddot{\phi}_0}L_{\dot{q}\dot{q}}-L_{\ddot{\phi}_0\dot{q}}^2=0,
\end{align}
of the Hessian
\begin{align}
\bm{A}\equiv 
\begin{pmatrix}
L_{\ddot{\phi}_0\ddot{\phi}_0} & L_{\ddot{\phi}_0\dot{q}} \\
L_{\ddot{\phi}_0\dot{q}} &
L_{\dot{q}\dot{q}}
\end{pmatrix}
.
\end{align}
Under this condition, \eqref{primary_c} turns to be a primary constraint on $p$. It can be easily confirmed that the preservation of the primary constraint through time evolution yields a secondary constraint which is linear in $p_{\phi}$ with a coefficient $\frac{\partial} {\partial \lambda}(L_{\dot{q}}|_{\phi_2=\phi_{\rm sol}})=\left. L_{\dot{q}\phi_2}\frac{\partial \phi_2}{\partial \lambda} \right|_{\phi_2=\phi_{\rm sol}} \neq 0$ due to $L_{\dot{q}\phi_2 }\ne 0$ and \eqref{f3}. Therefore, the secondary constraint fixes $p_{\phi}$ in terms of other canonical variables and then the on-shell Hamiltonian is no longer linear in canonical variables. 

Thus, the degeneracy condition guarantees the existence of constraint, and it indeed removes the Ostrogradsky ghost. 
Note that the degeneracy condition or the existence of additional constraint itself is still not sufficient.  
One needs to check that the additional constraint indeed eliminates linear momentum term from the on-shell Hamiltonian.
While we considered the simplest case, a set of degeneracy conditions can be derived for Lagrangian with multiple variables with second-order derivatives~\cite{Motohashi:2016ftl}, and even for Lagrangian with arbitrary higher-order derivatives~\cite{Motohashi:2017eya,Motohashi:2018pxg}, under which the absence of the Ostrogradsky ghosts can be confirmed.

\section{Degenerate theory with solvable auxiliary variables}%%%%%%%%%%%%%%%%%%%%%%%%%%%%%%%%%%%%%%%%%
\label{sec_aux1}

Generalizing the argument in \S\ref{sec_deg}, let us proceed to consider the following Lagrangian with $\xi^i$:
\begin{align} \label{Lpqx}
L=L(\ddot{\phi}_0,\dot{\phi}_0,\phi_0,\dot{q},q,\xi^i).
\end{align}
In contrast to \eqref{L1}, $\xi^i$ are a priori auxiliary variables.
The Euler-Lagrange equations for $\xi^i$ are given by 
\begin{align}
L_i=0. \label{Lxi}
\end{align}
Under the assumption ${\rm det}(L_{ij}) \neq 0$, from the implicit function theorem, all variables $\xi^i$ can be algebraically solved in terms of $\ddot{\phi}_0,\dot{\phi}_0,\phi_0,\dot{q},q$.
Again, this system has in principle three degrees of freedom, and one of them is the Ostrogradsky ghost, which we can eliminate by imposing degeneracy condition as we shall see below.

\subsection{Degeneracy condition}%%%%%%%%%%%%%%%%%%%%%
\label{sec_aux11}

In this section, we derive degeneracy condition for the theory~\eqref{Lpqx} under the assumption ${\rm det} (L_{ij})\neq 0$, with which we can solve \eqref{Lxi} for all $\xi^i$.
We shall address the case with $L_{ij}= 0$, which means $\xi^i$ are Lagrange multipliers (and we shall denote them $\lambda^a$), in \S\ref{sec_aux2}.
A caveat here is that even if ${\rm det}(L_{ij})\neq 0$, there is an exceptional case where the auxiliary variable can be Lagrange multiplier by a redefinition of the auxiliary variable.  
A simple example is 
\be \label{lagmulex} L=f_1(\ddot{\phi}_0,\dot{\phi}_0,\phi_0,\dot{q},q) + \xi^2 f_2(\ddot{\phi}_0,\dot{\phi}_0,\phi_0,\dot{q},q) , \ee
where the equation of motion of $\xi$
admits two qualitatively different solutions
\begin{align} \label{xif2}
    \xi=0 ~~{\rm or}~~ f_2=0
    .
\end{align}
If one chooses the first branch $\xi=0$, the constraint $L_{\xi}=0$ is the equation to fix $\xi$.  On the other hand, for the second branch $f_2=0$, the constraint $L_{\xi}=0$ no longer determines $\xi$.  Instead, $L_{\xi}=0~(f_2=0)$ is a nonholonomic constraint on $\phi_0,q$. In this example, the property of the constraint $L_{\xi}=0$ depends on the branch of the solution and the resultant systems are completely different.
In this section, we choose the branch satisfying ${\rm det}(L_{ij})\not\approx 0$ so that one can solve the constraint equation~\eqref{Lxi} for $\xi$, where $\not\approx$ means that the equality does not hold after taking into account all the constraints and their branches.
In the above example, among the two branches~\eqref{xif2} we choose the first branch $\xi=0$, where $\xi$ is determined by the constraint equation.

We denote $\xi^i_{\rm sol}=\xi^i_{\rm sol}(\ddot{\phi}_0,\dot{\phi}_0,\phi_0,\dot{q},q)$ as the solutions of \eqref{Lxi}. Plugging the solution to \eqref{Lpqx}, the Lagrangian is formally given by
\begin{align}
\tL(\ddot{\phi}_0,\dot{\phi}_0,\phi_0,\dot{q},q)=L|_{\xi^i=\xi^i_{\rm sol}} .
\end{align}
The degeneracy condition of $\tL$ can be obtained by computing the determinant of the Hessian of the Lagrangian $\tL$.
A difficulty here is that it is generally hard (or may be impossible globally) to obtain the explicit solutions $\xi^i=\xi^i_{\rm sol}$.
However, one can obtain the degeneracy condition without obtaining the explicit solutions.
In parallel to \eqref{f1}--\eqref{f3},
plugging $\xi^i=\xi^i_{\rm sol}$ into \eqref{Lxi} and taking derivatives, 
we can write down derivatives of $\xi^i_{\rm sol}$ in terms of derivatives of the original Lagrangian as
\begin{align} \label{fxi}
\frac{\partial \xi_{\rm sol}^i}{\partial \dot{q}}=-\left. L_{\dot{q}j}L^{ij} \right|_{\xi^i=\xi^i_{\rm sol}}, \quad
\frac{\partial \xi_{\rm sol}^i}{\partial \ddot{\phi}_0}=-\left. L_{\ddot{\phi}_0 j}L^{ij} \right|_{\xi^i=\xi^i_{\rm sol}},
\end{align}
where $L^{ij}$ is the inverse matrix of $L_{ij}$.
The Hessian of the Lagrangian $\tL$ is then given by
\begin{align}
\bm{A}\equiv 
\begin{pmatrix}
\tL_{\ddot{\phi}_0\ddot{\phi}_0} & \tL_{\ddot{\phi}_0\dot{q}} \\
\tL_{\ddot{\phi}_0\dot{q}} & \tL_{\dot{q}\dot{q}}
\end{pmatrix}
=
\left.
\begin{pmatrix}
a & b \\
b & k
\end{pmatrix}
\right|_{\xi^i=\xi^i_{\rm sol}},
\end{align}
where using \eqref{Lxi} and \eqref{fxi} the elements are given by 
\begin{align}
a &=L_{\ddot{\phi}_0\ddot{\phi}_0}-L_{\ddot{\phi}_0i}L^{ij}L_{\ddot{\phi_0}j}, \\
b &=L_{\ddot{\phi}_0\dot{q}}-L_{\ddot{\phi}_0i}L^{ij}L_{\dot{q}j}, \\
k &=L_{\dot{q}\dot{q}}-L_{\dot{q}i}L^{ij}L_{\dot{q}j}.
\end{align}
The degeneracy condition is given by
\begin{align} \label{deg-0}
\left[ ak-b^2  \right]_{\xi^i=\xi^i_{\rm sol}} = 0 .
\end{align}
Taking into account the constraint equation~\eqref{Lxi} instead of $\xi^i=\xi^i_{\rm sol}$, the degeneracy condition~\eqref{deg-0} can be equivalently written as a useful form 
\begin{align}
ak-b^2  =f^i L_i . \label{deg}
\end{align}
Here, $f^i$ is a set of regular functions under $\xi^i=\xi^i_{\rm sol}$. Note that one does not need to explicitly solve $\xi^i=\xi^i_{\rm sol}$ to check the degeneracy condition~\eqref{deg}. To show the degeneracy of the system, one only needs to check the equality of the left-hand side of \eqref{deg} to a linear combination of the constraints $L_{i}$, i.e.\ to show the existence of the functions $f^i$ that is regular on the constraint hypersurface.\footnote{If the solutions of the constraints have several branches, i.e.~the case that several constraint hypersurfaces exist in the phase space, one should choose one constraint hypersurface to evaluate the regularity of $f_i$. There can be the case that one branch satisfies \eqref{deg} but another does not, e.g.~$L=\frac{\ddot{\phi}_0^3}{6}+\frac{\xi^3}{3}-\frac{\xi^2 \ddot{\phi}_0}{2} +\frac{\dot{q}^2}{2}$. In this example, $f=\frac{\ddot{\phi}-\xi}{\xi(2\xi-\ddot{\phi})}$ is singular on the branch $\xi=0~(\ddot{\phi}\neq 0)$ but is regular on the correct branch $\xi-\ddot{\phi}=0$.}

\subsection{Examples}%%%%%%%%%%%%%%%%%%%%%
\label{sec_aux12}

As an application of the degeneracy condition~\eqref{deg}, let us see several examples.
As a consistency check, we begin with the simplest example 
\be \label{ex1} L = \f{\xi^2}{2} - \xi q + L_0(\ddot{\phi}_0,\dot{\phi}_0,\phi_0,\dot{q},q) .  \ee
Clearly, $\xi$ does not affect the degeneracy of the theory nor the dynamics of other variables, and $\xi=q$ from the Euler-Lagrange equation. 
As expected, in this case, \eqref{deg} is reduced to the standard degeneracy condition~\eqref{st-deg}. More generally, the degeneracy condition~\eqref{deg} is the same as the standard one~\eqref{st-deg} if the auxiliary variables couple with neither $\ddot{\phi}_0$ nor $\dot{q}$, i.e.\ if the Lagrangian takes the following form:
\be L =  L_0(\ddot{\phi}_0,\dot{\phi}_0,\phi_0,\dot{q},q) +L_1(\dot{\phi}_0,\phi_0,q,\xi^i).  \ee

Next example is 
\be \label{ex2} L=\f{\ddot\phi_0^2}{2} + \f{\dot q^2}{2} + \f{\xi^2}{2} - \xi\ddot\phi_0 + L_0(\dot{\phi}_0,\phi_0,q), \ee
which is similar to the toy model presented in \cite{Chen:2012au}.
We can check the degeneracy of the Lagrangian~\eqref{ex2} in several ways.
Since the Euler-Lagrange equation for $\xi$ is given by $\xi=\ddot\phi_0$, plugging it back into the Lagrangian we obtain $\tL=\dot q^2/2+ L_0(\dot{\phi}_0,\phi_0,q)$, which is clearly free from ghost.
Another way is to redefine the auxiliary variable as $\Xi\equiv \xi-\ddot{\phi}_0$. One then obtains the trivially ghost-free Lagrangian $L=\dot q^2/2+\Xi^2/2+L_0(\dot{\phi}_0,\phi_0,q)$. 
We can also apply the known degeneracy condition for multiple variables~\cite{Motohashi:2016ftl} by rewriting $- \xi\ddot\phi_0$ to $\dot\xi\dot\phi_0$ by integration by parts and regarding the Lagrangian as $L=L(\ddot\phi_0,\dot\phi_0,\dot q, q, \dot \xi,\xi)$.
Finally we can easily check that the Lagrangian~\eqref{ex2} satisfies the degeneracy condition~\eqref{deg}.

Another example is 
\begin{align} \label{ex3}
L=f_1(\xi)\ddot{\phi}_0+f_2(\xi)\dot{q}+L_0(\dot{\phi}_0,\phi_0,q, \xi)
.
\end{align}
with the assumption $L_{0qq}\neq 0$,
which is a generalization of the toy model of \cite{Gabadadze:2012tr}.
This example is clearly free from the Ostrogradsky ghost because the Lagrangian can be rewritten as the form $L=L(\dot{\phi}_0,\phi_0,q,\dot{\xi},\xi)$ when performing integration by parts.
One can then eliminate $q$ by the use of its Euler-Lagrange equation as far as $L_{0qq}\neq 0$ and obtain the Lagrangian in terms of two dynamical variables $\phi_0,\xi$ without higher derivatives\footnote{In the case $L_{0qq}=0$, the variable $q$ turns to be a Lagrange multiplier after the integration by parts. The system with Lagrange multipliers will be discussed in the next section.}. 
However, if we regard $\xi$ as an auxiliary variable, in general its Euler-Lagrange equation $L_\xi=0$ may not be explicitly solved. Nevertheless, the degeneracy condition~\eqref{deg} allows us to check the degeneracy without solving and substituting $\xi$.

As the final example, let us consider
\begin{align} \label{ex4}
    L&=\ddot{\phi}_0^3+(1+3\xi+3\dot{q})\ddot{\phi}_0^2+[\xi+3\xi^2+3(1+2\xi)\dot{q}+3\dot{q}^2]\ddot{\phi}_0
    \nn
    &~~~+\dot{q}^3+(2c+3\xi)\dot{q}^2+(2\xi+3\xi^2)\dot{q}+\xi^3 +L_0(\dot{\phi}_0,\phi_0,q)
    ,
\end{align}
where $c$ is a constant and $L_{0qq}\neq 0$. A priori, it might not be so straightforward to derive the Ostrogradsky ghost-free condition without using \eqref{deg}. On the other hand, by the use of \eqref{deg}, one can easily obtain the degeneracy condition
\begin{align}
    ak-b^2=\frac{2(c-1)}{3(\xi+\dot{q}+\ddot{\phi}_0)}=0.
\end{align}
Therefore, the Lagrangian \eqref{ex4} is free from the Ostrogradsky ghost if $c=1$. 
Indeed, \eqref{ex4} with $c=1$ can be reduced to the form of \eqref{ex3} by redefining the auxiliary variable. The Lagrangian \eqref{ex4} can be rewritten as
\begin{align}
    L=(\ddot{\phi}_0+\dot{q}+\xi)^3+2\dot{q}(\ddot{\phi}_0+c \dot{q}+\xi)+\ddot{\phi}_0 (\ddot{\phi}_0+c \dot{q}+\xi)+L_0(\dot{\phi}_0,\phi_0,q)
\end{align}
Therefore, for $c=1$, we can further rewrite it as
\begin{align}
    L=\Xi^3+2\dot{q}\Xi+\ddot{\phi}_0 \Xi + L_0(\dot{\phi}_0,\phi_0,q),
\end{align}
where we have redefined the auxiliary variable $\Xi\equiv \ddot{\phi}_0+\dot{q}+\xi$. In parallel to \eqref{ex3}, we can perform integration by parts to remove
$\dot{q},\ddot{\phi}_0$ from the Lagrangian and can eliminate $q$ by using its equation of motion; then, we conclude the Lagrangian no longer has a higher derivative nor auxiliary variable.

Let us summarize general lessons from these examples. 
For simple models there are several ways to check the degeneracy in general.
While some of them could be more straightforward than checking the degeneracy condition~\eqref{deg}, they are case-by-case basis whereas the latter always applies and hence provides a systematic check. 
Previously known ghost-free examples can be also understood by the degeneracy condition~\eqref{deg} from a unified point of view.
Furthermore, the degeneracy condition \eqref{deg} is a powerful tool especially to investigate more involved Lagrangians, and helps to extract a case free from the Ostrogradsky ghost.

\section{Theories with Lagrange multipliers}%%%%%%%%%%%%%%%%%%%%%%%%%%%%%%%%%%%%%%%%%
\label{sec_aux2}

In this section, we shall consider theories with Lagrange multipliers $\lambda^a$ as well as solvable auxiliary variables $\xi^i$ we considered in \S\ref{sec_aux1}. The inclusion of the Lagrange multipliers and the constraints implemented by $\lambda^a$ enables us to discuss various systems in a unified way: higher-derivative Lagrangian can be reduced to a Lagrangian with at most first-order derivatives (see Eqs.~\eqref{phi_eq1} and \eqref{phi_eq2}), and a system with a first class constraint can be reduced a system with a couple of second class constraints by introducing a gauge fixing condition as a constraint. We therefore do not consider either systems with higher derivatives or with first class constraints in this section. 
Note that in general the Hamiltonian is not necessarily linear in $\lambda^a$ even when the Lagrangian is linear in the Lagrange multipliers, and vice versa. 
In the present case, we assume that the Lagrangian is linear in $\lambda^a$ and nonlinear in $\xi^i$, but it is not necessarily the case for the Hamiltonian.

Below we shall address three types of Lagrangians, which are summarized in Table~\ref{tab_1}.
We consider system with holonomic constraints in \S\ref{sec_aux21} and system with nonholonomic constraints in \S\ref{sec_aux22}, in which we find that the Ostrogradsky-like ghost degrees of freedom show up due to the existence of constraints.  
As mentioned above, an example of nonholonomic system~\eqref{phi_eq1} is related to a higher-derivative system, and hence one can interpret the ghosts in this system as a consequence of either higher derivatives or nonholonomic constraints.  However, as we shall see below, there exist pathological nonholonomic systems which are no longer equivalent to higher derivative ones.  In this case, the Ostrogradsky theorem does not apply. Therefore, ghosts associated with nonholonomic constraints are a generalization of the Ostrogradsky ghosts. 
We argue how to eliminate these ghosts by employing $\xi^i$ in \S\ref{sec_aux23}, and provide a ghost-free criterion. In \S\ref{sec_aux24}, we briefly discuss more involved systems than the systems studied in \S\ref{sec_aux21}--\S\ref{sec_aux23}.

\subsection{Holonomic constraints}%%%%%%%%%%%%%%%%%%%%%
\label{sec_aux21}

Before discussing a system with nonholonomic constraints including derivatives, let us consider an elementary system:
\begin{align} \label{Lholo}
    L=L_0(\dot{q}^I,q^I)+\lambda^a C_a(q^I)
\end{align}
which has the $m$ holonomic constraints
\begin{align}
    C_a(q^I)=0, \label{holo}
\end{align}
between variables $q^I$ but not their derivatives.
Here we assume that the constraints are linearly independent.

Basically, if one can globally solve the set of holonomic constraints~\eqref{holo} for $m$ components of the variables $q^I$ in terms of others, one can substitute them back into the Lagrangian~\eqref{Lholo} and obtain a Lagrangian without constraints.  It manifests that the phase space dimension is reduced by the holonomic constraints.  
Below, to highlight the difference from the system with nonholonomic constraints which we shall consider in \S\ref{sec_aux22}, we consider more general process, which also applies to the case where the holonomic constraints are solvable only locally.

The total Hamiltonian corresponding to \eqref{Lholo} is
\begin{align} \label{Htotholo}
    H_{\rm tot}=\dot{q}^I p_I-L_0(\dot{q}^I,q^I)-\lambda^a C_a(q^I) +\zeta^a \pi_a,
\end{align}
where $\dot{q}^I=\dot{q}^I(q^J,p_J)$ are understood as the solutions of the equations
\begin{align}
    p_I = L_{0I}(\dot{q}^J,q^J).
\end{align}
It is worth emphasizing that the Hamiltonian is linear in $\lambda^a$, i.e.\ $\lambda^a$ are also the Lagrange multipliers even in the Hamiltonian language. As we will see in \S\ref{sec_aux22}, this is not the case when the constraints $C_a=0$ are nonholonomic.

The consistency conditions of the primary constraints $\pi_a\approx 0$ lead to secondary constraints
\begin{align}
    C_a(q^I) \approx 0.
\end{align}
The consistency of conditions for $C_a \approx 0$ give
\begin{align}
    D_a\equiv \{ C_a,H_{\rm tot}\} = \frac{\partial C_{a}}{\partial q^I} \dot{q}^I(q^J,p_J) \approx 0.
\end{align}
Further, the consistency conditions for $D_a\approx 0$ yield $E_a\equiv \{ D_a,H_{\rm tot}\}\approx 0$, which determine $\lambda^a$ in terms of $q^I,p_I$ if $\det \{C_a, D_b\} 
\approx \det \left( \frac{\partial C_a}{\partial q^I}L_0^{IJ} \frac{\partial C_b}{\partial q^J}  \right)  \not\approx 0$.  The consistency conditions of $E_a$ finally fix $\zeta^a$ and there are no further constraints.
In this case, we have $4m$ constraints on $(q^I,p_I,\lambda^a,\pi_a)$ where $\pi_a\approx 0, E_a\approx 0$ fix $(\lambda^a,\pi_a)$ and $C_a \approx 0,D_a \approx 0$ are constraints on $(q^I,p_I)$. Hence, the phase space dimension is $2(n-m)$. This implies nothing but that the holonomic constraint $C_a =0$ determines $m$ components of the variables $q^I$ in terms of others.

If $\det \{C_a, D_b\} \approx 0$, some of $\lambda^a$ are not determined by $E_a \approx 0$. Since we have considered a system without first class constraints, all Lagrange multipliers $\zeta^a$ must be determined. However, the consistency conditions of $E_a \approx 0$ do not fix all $\zeta^a$; instead, some of linear combinations of $E_a \approx 0$ must be the constraints on the variables $(\lambda^a,q^I,p_I)$. We should continue to check the consistency conditions of these constraints until all $\zeta^a$ are fixed. In this case, the number of the constraints is larger than $2m$ and then the phase space dimension is smaller than $2(n-m)$.

\subsection{Nonholonomic constraints}%%%%%%%%%%%%%%%%%%%%%
\label{sec_aux22}

Let us proceed to consider the Lagrangian with nonholonomic constraints
\begin{align}\label{Lnonholo}
    L=L_0(\dot{q}^I,q^I)+\lambda^a C_a(\dot{q}^I,q^I) ,
\end{align}
under the nondegeneracy condition $\det (L_{IJ}) \not\approx 0$.
An important nature of the nonholonomic constraints is that in general the constraints $C_a=0$ do not reduce the phase space dimension of $(q^I,p_I)$, which is precisely the origin of the Ostrogradsky-like ghosts as we shall show below.

Under the nondegeneracy condition, the total Hamiltonian is
\begin{align}\label{Htotnon}
 H_{\rm tot}=\dot{q}^I p_I-L_0(\dot{q}^I,q^I)-\lambda^a C_a(\dot{q}^I,q^I) +\zeta^a \pi_a,
\end{align}
where $\dot{q}^I$ are given by solutions of algebraic equations
\begin{align} \label{pIdef}
    p_I=L_{0I}(\dot{q}^J,q^J)+\lambda^a C_{aI}(\dot{q}^J,q^J). 
\end{align}
Hence, the functions $\dot{q}^I$ are now functions of not only $(q^I,p_I)$ but also $\lambda^a$. 
From \eqref{pIdef}, we obtain
\be \label{dqlam} \f{\pa\dot q^I}{\pa\lambda^a} = - L^{IJ}C_{aJ}. \ee
As a result, the total Hamiltonian~\eqref{Htotnon} is generally nonlinear in $\lambda^a$, which is a crucial difference from the holonomic system we addressed in \S\ref{sec_aux21} (see \eqref{Htotholo}).
This difference can be captured by a matrix 
\begin{align}\label{defMab}
    M_{ab}\equiv \frac{\partial^2 H_{\rm tot}}{\partial \lambda^a \partial \lambda^b}
    =-\frac{\partial C_a}{\partial \lambda^b}
    =C_{aI}L^{IJ}C_{bJ},
\end{align}
where we used \eqref{pIdef} and \eqref{dqlam}.  
For the holonomic system, $M_{ab}$ vanishes identically. 
In contrast, the nonholonomic system allows nonvanishing $M_{ab}$.
To extract the essence of the nonholonomic system, 
we assume the full nondegeneracy, 
\begin{align}
    {\rm det}(M_{ab})\not\approx 0. \label{detM}
\end{align}
To satisfy the condition~\eqref{detM}, it is necessary that ${\rm rank}\, C_{aI}=m$ under $C_a=0$, i.e.\ the nonholonomic constraints $C_a\approx 0$ are linearly independent with respect to $\dot q^I$.
If some of $C_a\approx 0$ are linearly dependent with respect to $\dot q^I$, one can eliminate $\dot q^I$ from such constraints, which implies that they are actually holonomic constraints.
Requiring \eqref{detM}, we focus on the case where all $C_a\approx 0$ are essentially nonholonomic constraints.

For later convenience, without loss of generality, we suppose that $C_{aI}$ satisfies a condition $\det \bm{C} \ne 0$ under $C_a=0$ with a $n\times n$ matrix $\bm{C}$ defined as 
\begin{align}
    \bm{C}&\equiv 
\begin{pmatrix}
   C_{aI} \\ \cline{1-1} 
   \begin{matrix} \, \bm{0}_{n-m,m}  & \vline\,\, \bm{1}_{n-m} \end{matrix} 
\end{pmatrix} ,
 \label{mat_C11}
\end{align}
where $\bm{0}_{i,j}$ and $\bm{1}_{i}$ denote the $i\times j$ zero matrices and the $i\times i$ identity matrices, respectively.
Let us explain this condition more explicitly.
As mentioned above, ${\rm rank}\, C_{aI}=m$ under $C_a=0$ is the necessary condition to satisfy our assumption \eqref{detM}. 
Simple examples of such $C_{aI}$ are 
\be
\begin{pmatrix}
    \bm{1}_{m} & \vline\,\, \bm{0}_{m,n-m}
\end{pmatrix} ,
\quad
\begin{pmatrix}
    \bm{0}_{m,n-m} & \vline\,\, \bm{1}_{m}
\end{pmatrix} .
\ee
The former satisfies the condition $\det \bm{C} \ne 0$ but the latter is not. 
Of course, the two matrices are equivalent under relabelling of the column index $a$.
Hence, even if $C_{aI}$ takes the latter form, we can relabel $a$ and make $C_{aI}$ to satisfy the condition $\det \bm{C} \ne 0$.
More generally, so long as ${\rm rank}\, C_{aI}=m$, we can always relabel $a$ and $I$ for $C_{aI}$ by redefinition of the basis so that $\det \bm{C} \ne 0$ is satisfied.

With the above assumptions, let us proceed to the Hamltonian analysis.
The consistency condition of the primary constraints
\begin{align}
    \pi_a \approx 0 , \label{pri}
\end{align}
yield the secondary constraints
\begin{align}
    C_a(\dot{q}^I,q^I)\approx 0 ,
\end{align}
where we recall that $\dot{q}^I$ are functions of $(q^J,p_J,\lambda^a)$. Due to the assumption ${\rm det}(M_{ab}) = -\det \frac{\partial C_a}{\partial \lambda^b} \not\approx 0$, from the implicit function theorem, the secondary constraints $C_a\approx 0$ can be solved for $\lambda^a$ and fix them in terms of $(q^I,p_I)$, 
\begin{align}
    \lambda^a\approx F^a(q^I,p_I). \label{sec}
\end{align}
Requiring the time preservation of the secondary constraints $C_a\approx 0$ yields the tertiary constraints $D_a\equiv \{ C_a,H_{\rm tot}\}\approx 0$, which fix the Lagrange multipliers $\zeta^a$ due to the $\lambda^a$ dependence of $\dot q^I$,
and hence there are no further constraints. 
Therefore, in contrast to the previous case with holonomic constraints, in the present case we have only $2m$ constraints on $(q^I,p_I,\lambda^a,\pi_a)$: $\pi_a\approx 0, C_a\approx 0$ fix $(\lambda^a,\pi_a)$, and all $(q^I,p_I)$ remain unconstrained. The phase space dimension is thus $2n$. The assumption ${\rm det}(M_{ab})\not\approx 0$ is the condition that the constraints $C_a\approx 0$ do not reduce the phase space dimension of the dynamical variables $q^I$.

We then discuss the non/existence of the local extremum of the on-shell Hamiltonian
\begin{align}\label{Honshell}
    H_{\text{on-shell}}(q^I,p_I) \equiv \dot{q}^I_{\lambda=F} p_I-L_0(\dot{q}^I_{\lambda=F},q^I)
\end{align}
by following the second partial derivative test
where $\dot{q}^I_{\lambda=F}=\dot{q}^I(q^J,p_J,\lambda)|_{\lambda=F}$ are now functions of only $(q^J,p_J)$ because we have substituted the solution \eqref{sec}. 
The on-shell Hamiltonian is a function of the independent variables $(q^I,p_I)$.

First, we identify the stationary points, at which the first derivative vanishes: 
\begin{align}
\frac{\partial H_{\text{on-shell}}}{\partial q^I}=\frac{\partial H_{\text{on-shell}}}{\partial p_I}=0. \label{stationary}
\end{align}
Second, we check the eigenvalue of Hessian matrix at the stationary points.
In the present case, the $2n\times 2n$ Hessian matrix of the on-shell Hamiltonian is given by
\begin{align}
    \bm{H}=
    \begin{pmatrix}
    \bm{H}_{11} & \bm{H}_{12} \\
    \bm{H}_{12}^T & \bm{H}_{22}
    \end{pmatrix}
\end{align}
where $n\times n$ sub-matrices are defined by
\begin{align}
    (\bm{H}_{11})^{IJ} &\equiv \frac{\partial^2 H_{\text{on-shell}}}{\partial p_I \partial p_J}
    , \label{H11} \\
    (\bm{H}_{12})^I{}_J &\equiv \frac{\partial^2 H_{\text{on-shell}}}{\partial p_I \partial q^J}
    , \\
    (\bm{H}_{22})_{IJ} &\equiv \frac{\partial^2 H_{\text{on-shell}}}{\partial q^I \partial q^J}
    .
\end{align}
If the Hessian matrix is positive or negative definite at a stationary point, i.e.\ all the eigenvalues are positive or negative, then the stationary point is a local minimum or maximum, respectively.
On the other hand, if the Hessian is indefinite at a stationary point, i.e.\ it has both positive and negative eigenvalues, the stationary point is a saddle point. 
Otherwise, the second derivative test is inconclusive.
For instance, if the Hessian is semi-definite including vanishing eigenvalues at a stationary point, it can be a local extremum or saddle point.

Hence, the absence of local extremum can be proved if there is no stationary point or if the Hessian $\bm{H}$ is an indefinite matrix 
at any stationary points, i.e.\ all stationary points are saddle points.

Let us first compute the first derivative of the on-shell Hamiltonian~\eqref{Honshell}.
Since $C_a=0$ is identically satisfied when we substitute $\lambda^a=F^a$, taking derivatives with respect to $p_I$ and $q^I$ yields identities
\begin{align}
   \label{cpi} C_{aJ}^{\lambda=F}\frac{\partial \dot{q}^J_{\lambda=F}}{\partial p_I} &=0\,, \\
   \label{cqi} \left. \frac{\partial C_{a}}{\partial q^I}\right|_{\lambda=F}+ C_{aJ}^{\lambda=F}\frac{\partial \dot{q}^J_{\lambda=F}}{\partial q^I}&=0.
\end{align}
We then have
\begin{align}
    \frac{\partial H_{\text{on-shell}}}{\partial p_I}&=[p_J-L_{0J}]_{\lambda=F} \frac{\partial \dot{q}^J_{\lambda=F}}{\partial p_I} +\dot{q}^I_{\lambda=F}
    \nn
    &=F^aC_{aJ}^{\lambda=F}\frac{\partial \dot{q}^J_{\lambda=F}}{\partial p_I} +\dot{q}^I_{\lambda=F}
    \nn
    &=\dot{q}^I_{\lambda=F} ,
\end{align}
where we used \eqref{pIdef} and \eqref{cpi}.
Likewise, using \eqref{pIdef} and \eqref{cqi} we obtain 
\begin{align}
    \frac{\partial H_{\text{on-shell}}}{\partial q^I} &=[p_J-L_{0J}]_{\lambda=F} \frac{\partial \dot{q}^J_{\lambda=F}}{\partial q^I} -\left.\f{\pa L_0}{\pa q^I}\right|_{\lambda=F}
    \nn
    &=F^aC_{aJ}^{\lambda=F}\frac{\partial \dot{q}^J_{\lambda=F}}{\partial q^I} -\left.\f{\pa L_0}{\pa q^I}\right|_{\lambda=F} 
    \nn
    &=-F^a\left. \frac{\partial C_{a}}{\partial q^I}\right|_{\lambda=F} -\left.\f{\pa L_0}{\pa q^I}\right|_{\lambda=F} 
    \nn    
    &= - \left. \f{\pa L}{\pa q^I} \right|_{\lambda=F} .
\end{align}  
Then, the stationary condition~\eqref{stationary} reads 
\be \dot{q}^I_{\lambda=F} = \left. \f{\pa L}{\pa q^I} \right|_{\lambda=F} = 0. \label{stationary2} \ee
Therefore, if $\dot{q}^I=\f{\pa L}{\pa q^I} =0$ under $C_a=0$ cannot be a solution of the system, the on-shell Hamiltonian has no stationary point and then no local extremum. 
A simple example is a constraint $\dot{q}-c=0$ with a nonzero constant $c$, which does not allow $\dot q\approx 0$.
In this case, regardless of the shape of the potential, $q$ cannot stop and should continue to move with nonzero velocity, developing an instability.
Another example is a constraint $\dot{q}=0$ with a linear potential, for which $\f{\pa L}{\pa q}\approx 0$ is not allowed.
Again, the requirements on the dynamics of $q$ from the constraint and the potential are incompatible.
We remark that the instabilities in these examples are not equivalent to the pathological instability originated from higher derivatives.

Next, let us assume that the Hamiltonian has a stationary point, and analyze the signature of the Hessian $\bm{H}$.
For each stationary point, we can multiply $\pm 1$ to redefine $\bm{H}$ so that it is not negative definite.  
Therefore, we can assume that $\bm{H}$ is either positive semi-definite or indefinite at stationary points without loss of generality.
The upper-left component~\eqref{H11} of the Hessian can be written as 
\begin{align}
    (\bm{H}_{11})^{IJ}=\frac{\partial \dot{q}^I_{\lambda=F}}{\partial p_J},
\end{align}
which from \eqref{cpi} has $m$ zero eigenvalues associated with the kernel $C_{aI}^{\lambda=F}$.
In general, the signature of a symmetric matrix is unchanged under the transformation
\begin{align}
    \bm{H} \rightarrow \bm{H}'=\bm{P}^{T}\bm{H}\bm{P},
\end{align}
where $\bm{P}$ is a nondegenerate matrix. 
In the present case, we choose $\bm{P}$ as 
\begin{align}
\bm{P}&=
\begin{pmatrix}
   \bm{C}_{\lambda=F} & \bm{0}_{n,n} \\ 
   \bm{0}_{n,n} & \bm{1}_{n}
\end{pmatrix},
\end{align}
where $\bm{C}_{\lambda=F}$ is the $n\times n$ nondegenerate matrix defined in \eqref{mat_C11} with \eqref{sec}.
Using \eqref{cpi} and \eqref{cqi}, we obtain the form
\begin{align}
    \bm{H}'&=
\begin{pmatrix}
   \bm{0}_{m, m} & \begin{matrix} \bm{0}_{m, n-m} & -\bm{C_q} \end{matrix} \\
   \begin{matrix} 
   \bm{0}_{n-m, m} \\ 
   -\bm{C_q}^T 
   \end{matrix}& \ddots 
\end{pmatrix},
\label{block_H}
\end{align}
where
\begin{align}
     (\bm{C_q})_{aI}&\equiv -C_{aI}^{\lambda=F} \frac{\partial^2 H_{\text{on-shell}}}{\partial p_I \partial q^J} =\left. \frac{\partial C_a}{\partial q^I}\right|_{\lambda=F}, \label{defCq}
\end{align} 
and the dotted components $\ddots$ in $\bm{H}'$ are the same as the corresponding components of $\bm{H}$, which we shall show are irrelevant to the following calculations.

In general, if a symmetric matrix is positive semi-definite, all principal minors of the symmetric matrix are non-negative (see e.g.\ Ref.~\cite{Meyer:2010} \S7.6 Positive Definite Matrices, page 566). 
The contraposition of this theorem tells us:  If there exists a negative principal minor for a symmetric matrix, the matrix is not positive semi-definite.
Recalling that in the present case we have assumed that the Hessian is positive semi-definite or indefinite at the stationary points, we can conclude that the Hessian $\bm{H}$ is indefinite if there exists a negative principal minor. 
We can show that there exists a negative principal minor if the condition 
\begin{align}
    \bm{C_q} \neq 0, \label{Cqnz}
\end{align}
is satisfied as follows.
Under the condition~\eqref{Cqnz}, there exists a nonzero $a_i$-$I_j$ component of $\bm{C_q}$, denoted by $(\bm{C_q})_{a_i I_j}$. The matrix $\bm{H}'$ clearly has the following negative principal minor of order two,
\begin{align}
    \begin{vmatrix}
     0 & -(\bm{C_q})_{a_i I_j} \\
     -(\bm{C_q})_{a_i I_j} & (\bm{H}_{22})_{I_j I_j} 
    \end{vmatrix}
    <0,
\end{align}
regardless of the value $(\bm{H}_{22})_{I_j I_j}$, which is the $I_j$-$I_j$ component of $\bm{H}_{22}$.
If \eqref{Cqnz} is satisfied at all the stationary point, we conclude that the Hessian $\bm{H}$ is indefinite and then the on-shell Hamiltonian only has a saddle point at a stationary point.
The existence of the negative eigenvalue implies that there exists a ghostly degree of freedom.
Therefore, the final criterion is the condition~\eqref{Cqnz} at any stationary points.

In summary, 
the absence of local extremum
of the on-shell Hamiltonian~\eqref{Honshell} for the Lagrangian~\eqref{Lnonholo} with the nonholonomic constraints is proved if the following conditions are satisfied:\footnote{Note that there is a possibility that the Hamiltonian takes a finite value at the boundary of the domain of $(q^I,p_I)$ and then the Hamiltonian is bounded. We will revisit this point in \S\ref{sec_canonical}.
}
\begin{enumerate}
\item \label{con1} 
${\rm det}(L_{IJ})\not\approx 0$: The system is nondegenerate with respect to all variables included in $L_0$.
\item \label{con2} 
${\rm det}(M_{ab}) \not\approx 0$: The nonholonomic constraints do not reduce the phase space dimension of the all variables included in $L_0$. 
\item \label{con3} 
Either of the following two cases:
\begin{enumerate}
\item[a.] $\dot{q}^I\approx 0 \approx \f{\pa L}{\pa q^I}$ is not allowed: Then, $H_{\text{on-shell}}$ does not have stationary points.
\item[b.] $\bm{C_q} \neq 0$ at any stationary points: Then, $H_{\text{on-shell}}$ has saddle points only and no local extremum.
\end{enumerate} 
\end{enumerate}
For a given Lagrangian with nonholonomic constraints, the conditions~\ref{con1} and \ref{con2} are straightforwardly checked by computing $\det(L_{IJ})$ and $\det(M_{ab})$. 
As we considered simple examples below \eqref{stationary2}, the condition~\ref{con3}-a captures the instabilities in theories which are not related to a higher-derivative theory.
This case is qualitatively different from the pathological Ostrogradsky ghost. On the other hand, theories satisfying the condition~\ref{con3}-b (and the conditions~\ref{con1} and \ref{con2}) possess ghostly degree of freedom.
One may think that the condition~\ref{con3}-b would suggest that the pathological theories are equivalent to higher derivative systems at least locally. However, it is not always the case. For example, if a constraint takes a form $\delta \dot{ q}^1+\delta q^2 \simeq 0$ around a stationary point where $\delta q^1$ and $\delta q^2$ are perturbations around the stationary point, one can solve the constraint as $\delta q^2\simeq -\delta\dot{  q}^1$ and then conclude the system is locally equivalent to a higher derivative system. On the other hand, a system with a constraint $\delta \dot{q}^1+\delta q^1\simeq 0$ is not a higher derivative one even locally. Rather, this is a lower derivative system but it shares the same pathology as the Ostrogradsky system (see \eqref{H_uns} with $b_0=0$ below).
For more general case, the conditions~\ref{con3}-a and \ref{con3}-b can be checked at least numerically since the stationary points of the on-shell Hamiltonian satisfy $\dot{q}^I=0$ under which the equations of motion and the constraints are algebraically solved, and one can check if there exists any inconsistency.

This set of conditions applies to the general Lagrangian~\eqref{Lnonholo} with nonholonomic constraints, which includes higher-derivative theories as a subclass. 
Hence, it is a natural generalization of the Ostrogradsky theorem.
Indeed, we shall see in \S\ref{sec_ex2} that for higher-derivatives theories only the condition~\ref{con1} is relevant, which is precisely related to the nondegeneracy assumption of the Ostrogradsky theorem, and the conditions~\ref{con2} and \ref{con3}-b are automatically satisfied. 
In this case, the on-shell Hamiltonian~\eqref{Honshell} exhibits the linear dependence on canonical momenta.
However, in general it is not always linear in canonical momenta since $\dot{q}^I_{\lambda=F}$ are functions of $(q^J,p_J)$.
Even for such more general cases, the conditions~\ref{con1}--\ref{con3} are robust. 
If they are satisfied, the Hamiltonian of the system does not admit a local extremum. 
In particular, when the condition \ref{con3}-b is satisfied, we have proved that the Hessian admits a negative eigenvalue at the stationary point implying the existence of a ghostly degree of freedom.
In addition to the simple examples considered below \eqref{stationary2}, we shall see several examples in \S\ref{sec_aux24} and \S\ref{sec_ex1}, for which the Ostrogradsky theorem does not apply but our criteria can detect the pathology of the Hamiltonian.

\subsection{Ghost-free Lagrangian with nonholonomic constraints}%%%%%%%%%%%%%%%%%%%%%
\label{sec_aux23}

We then discuss how to evade the Ostrogradsky-like ghost found in \S\ref{sec_aux22}.
First, to violate the conditions~\ref{con1}--\ref{con3}, one may consider a possibility to convert the nonholonomic constraints of the Lagrangian~\eqref{Lnonholo} into holonomic constraints, by introducing nondynamical variables $\xi^i$.
Let us consider
\begin{align} \label{L_xi_lambda}
    L=L_0(\dot{q}^I,q^I,\xi^i)+\lambda^a C_a(\dot{q}^I,q^I,\xi^i).
\end{align}
If 
\begin{align} \label{gf_con1}
    {\rm rank}(C_{ai})=m ,
\end{align} 
the constraints $C_a=0$ can be solved for $m$ components of $\xi^i$ in terms of $\dot{q}^I,q^I$. 
The constraints $C_a=0$ can be now essentially interpreted as holonomic constraints to remove $m$ of $\xi^i$ from the phase space\footnote{ Precisely, $\xi^i$ are nondynamical and thus $\xi^i$ can be removed by solving their Euler-Lagrange equations (see, however, the discussions in \S\ref{sec_aux24} and \S\ref{sec_ex3}). }.

Strictly speaking, however, $C_a$ depend on derivatives $\dot q^I$ so one may wonder if $C_a=0$ are really interpreted as holonomic constraints without derivatives.  We can confirm it is indeed the case by a change of the variables which we shall demonstrate below.
Let us denote $\xi^a$ as the solvable $m$ components of $\xi^i$, and $\xi^{i'}$ as the remaining $\ell-m$ nondynamical variables. The constraints $C_a=0$ have been supposed to be solved as
\begin{align}
    \xi^a=\xi^a_{\rm sol}(\dot{q}^I,q^I,\xi^{i'}),
\end{align}
where $\xi^a_{\rm sol}$ are functions of $\dot{q}^I,q^I,\xi^{i'}$. After a redefinition of the Lagrange multipliers, the Lagrangian may be transformed into the form
\begin{align} \label{Lamb2}
    L=L_0(\dot{q}^I,q^I,\xi^i)+\lambda_a\left[ \xi^a-\xi^a_{\rm sol}(\dot{q}^I,q^I,\xi^{i'}) \right].
\end{align}
We then introduce new variables $\Xi^a$ via
\begin{align}
    \Xi^a=\xi^a-\xi^a_{\rm sol}(\dot{q}^I,q^I,\xi^{i'}).
\end{align}
As a result, the Lagrangian becomes
\begin{align}
    L=L_0(\dot{q}^I,q^I,\xi^{i'},\Xi^a )+\lambda_a \Xi^a,
\end{align}
which is clearly a system with the holonomic constraints $\Xi^a=0$. 
Following the prescription of \S\ref{sec_aux21}, we can erase $\Xi^a$ and obtain an equivalent Lagrangian
\begin{align} \label{L_xi_after}
    L=L_0(\dot{q}^I,q^I,\xi^{i'}).
\end{align}
Hence, with the condition~\eqref{gf_con1}, we can remove the constraints which the original Lagrangian~\eqref{L_xi_lambda} has.

However, \eqref{gf_con1} is a sufficient condition to remove the constraints $C_a=0$ but not a sufficient condition to obtain a healthy Hamiltonian.
We still need to check if the Lagrangian~\eqref{L_xi_after} leads to a healthy Hamiltonian or not.
Actually, the Lagrangian~\eqref{L_xi_after} can be linear in $\xi^{i'}$ even if the original Lagrangian~\eqref{L_xi_lambda} is nonlinear in $\xi^{i'}$. For instance, consider a system
\begin{align} \label{ex_ghost}
    L=\frac{1}{2}(\dot{q})^2+\frac{1}{2}(\xi^1+\xi^2)^2+(\xi^1+\xi^2)\xi^2+\lambda(\xi^1+\xi^2-q-\dot{q}).
\end{align}
For this system, $m=1$ and $C_{i}=(1,1)^T$, whose rank is $1$ and hence this model satisfies \eqref{gf_con1}.
Nevertheless, solving the constraint $\xi^1+\xi^2+q+\dot{q}=0$ for $\xi^1$ and substituting it, one obtains
\begin{align}
    L'=\frac{1}{2}(\dot{q})^2+\frac{1}{2}(\dot{q}+q)^2+(\dot{q}+q)\xi^2,
\end{align}
where $\xi^2$ turns to be a Lagrange multiplier and then has the Ostrogradsky-like ghost from Theorem~\ref{main_th}.

An improved condition to find a ghost-free theory is to require that the set $(\xi^i,\lambda^a)$ can be interpreted as a set of solvable variables.
In this case, $\lambda^a$ quit the role of the Lagrange multipliers and they are actually solvable auxiliary variables even if the Lagrangian linearly depends on them. 
Considering their Euler-Lagrange equations, the solvability condition is given by
\begin{align} \label{detD}
    \det \bm{D}\not\approx 0,
\end{align}
where
\begin{align}
\bm{D}\equiv 
    \begin{pmatrix}
       L_{ij} & C_{ai} \\
       C_{ai}^T & \bm{0}_m \\
    \end{pmatrix}
    .
\end{align}
It is straightforward to check that \eqref{ex_ghost} does not satisfy the condition~\eqref{detD}.
For general case, if the condition~\eqref{detD} is satisfied, after substituting the solutions of the Euler-Lagrange equations for $(\xi^i,\lambda^a)$, the Lagrangian $L'=L|_{\xi^i=\xi^i_{\rm sol},\lambda^a=\lambda^a_{\rm sol}}$ is a function of $q^I$ and $\dot{q}^I$ only. 
Unless some of $q^I$ are turned out to be auxiliary variables, ghosts associated with higher derivatives or constraints are absent.  
One can then proceed to the standard process such as checking the sign of the kinetic term, i.e.\ whether $L'_{IJ}$ is positive definite.

\subsection{On general Lagrangian}%%%%%%%%%%%%%%%%%%%%%
\label{sec_aux24}

Since the systems discussed in \S\ref{sec_aux21}--\S\ref{sec_aux23} are ideal cases, we briefly discuss involved systems in this subsection.
In \S\ref{sec_aux11} and \S\ref{sec_aux23} we have discussed the ambiguity associated with a redefinition of Lagrange multipliers (see the arguments on \eqref{lagmulex} and \eqref{Lamb2}).
For system with nonholonomic constraints, one can make use of this ambiguity to reformulate a degenerate Lagrangian to a nondegenerate Lagrangian, for the latter of which we can apply the conditions~\ref{con1}--\ref{con3}.
This technique also allows us to apply the conditions~\ref{con1}--\ref{con3} to a wider class of theories than the Ostrogradsky theorem.

The simplest toy model would be
\begin{align}
L_{\rm toy}=\lambda(\dot{q}-q). \label{L1_lam}
\end{align}
This Lagrangian is clearly free from higher derivative and degenerate as $L_{\dot q \dot q}=0$.  Hence, both the assumption of the Ostrogradsky theorem and the condition~\ref{con1} are not satisfied. 
Nevertheless, one can directly see that the corresponding Hamiltonian is unbounded.
There are two primary constraints
\begin{align}
    \pi \approx0 ,\quad p-\lambda \approx 0 ,
\end{align}
and no secondary constraints. The Hamiltonian on the constraint hypersurface is unbounded
\begin{align}
    H_{\text{on-shell}}=p q . \label{H_lam}
\end{align}

On the other hand, one can redefine the Lagrange multiplier $\lambda = \Lambda + \frac{1}{2}\dot{q}$ so that the Lagrangian becomes
\begin{align}
    L_{\rm toy}&=\frac{1}{2}\dot{q}^2-\frac{1}{2}\dot{q}q+\Lambda(\dot{q}-q) \nn
    &=\frac{1}{2}\dot{q}^2+\Lambda(\dot{q}-q) + {\rm total~derivative},
    \label{L1_Lam}
\end{align}
which is nondegenerate as $L_{\dot q \dot q}\ne 0$. 
In this form, the Lagrangian satisfies the conditions~\ref{con1}--\ref{con3}, and hence we can conclude that the Hamiltonian is pathological.
We can also directly check the Hamiltonian for the Lagrangian~\eqref{L1_Lam}.
There is only one primary constraint
\begin{align}
    \Pi \approx 0 ,
\end{align}
and its consistency condition yields the secondary constraint
\begin{align}
    \dot{q}-q\approx P-\Lambda-q \approx 0,
\end{align}
where the canonical momentum of $\Lambda$ and $q$ are respectively denoted as $\Pi$ and $P=\frac{\partial L_1}{\partial \dot{q}}= \dot{q}+\Lambda$.
The on-shell Hamiltonian is
\begin{align}
    H_{\text{on-shell}}=P q-\frac{1}{2}q^2 , \label{H_Lam}
\end{align}
which linearly depends on $P$ and has no local extremum.
The Hamiltonian \eqref{H_Lam} indeed coincides with \eqref{H_lam} because the momenta $p$ and $P$ are related by $P\approx p+\frac{1}{2}q$. 

Another involved system is a Lagrangian with both holonomic and nonholonomic constraints which can be interpreted as the violation of the condition \ref{con2}. In this case, it would be better to solve the holonomic constraints first and then apply the conditions~\ref{con1}--\ref{con3}. While in general the holonomic constraints may not be explicitly solved, one can solve the holonomic constraints at least locally and thus compute $L_{IJ},M_{ab},\bm{C_q}$ by the use of the implicit function theorem, in principle (see the computations in \S\ref{sec_aux1}). One can then check the conditions~\ref{con1}--\ref{con3} at the stationary points at which the equations of motion and the constraints are algebraically solved.

\section{Concrete examples}%%%%%%%%%%%%%%%%%%%%%%%%%%%%%%%%%%%%%%%%%
\label{sec_ex}

In this section we provide concrete examples for the  application of the general arguments developed in \S\ref{sec_aux2}.
In \S\ref{sec_ex1} we present a simple example to consider the cases considered in \S\ref{sec_aux21}, \ref{sec_aux22}, and \ref{sec_aux23} and highlight 
the pathology of the Hamiltonian. In the concrete example, we also find that the Hamiltonian is linear in the momentum and thus it exactly shows the same pathological behavior as the Ostrogradsky system even though the model cannot be rewritten to a higher-derivative theory.
In \S\ref{sec_ex2} we consider higher-derivative model, for which our criteria reduce to the Ostrogradsky theorem.
In \S\ref{sec_ex3} we revisit the healthy example in \S\ref{sec_over1} from the point of view of (non)holonomic system and the ghost-free criterion~\eqref{detD}, and discuss the interplay between the violation of condition~\ref{con1} or \ref{con2}. 
We address the violation of the condition~\ref{con3} in \S\ref{sec_ex4}.
We finally argue a possibility to obtain a bounded Hamiltonian from a Hamiltonian without a local minimum by restricting the domain of the variables in \S\ref{sec_canonical}.

\subsection{Ostrogradsky-like instability from nonholonomic constraints}%%%%%%%%%%%%%%%%%%%%%
\label{sec_ex1}

To show an explicit example of the result of \S\ref{sec_aux21}--\ref{sec_aux23}, we study a toy model
\begin{align} \label{Ltoy}
    L=\frac{a_1}{2}\dot{q}^2-\frac{a_2}{2}q^2-\frac{a_3}{2}\xi^2+\lambda(b_0+b_1 \dot{q} + b_2 q+b_3\xi).
\end{align}
with $a_1 \neq 0$.
The canonical momenta for $q,\lambda,\xi$ are
\begin{align}
   p=a_1 \dot{q}+b_2 \lambda, \quad 
   \pi = 0, \quad \varpi = 0,
\end{align}
the last two of which are primary constraints.
The total Hamiltonian is then given by
\begin{align}
    H_{\rm tot}=\frac{1}{2a_1}p^2+\frac{a_2}{2}q^2+\frac{a_3}{2}\xi^2-\lambda\left(b_0+\frac{b_1}{a_1}p+b_2 q+ b_3 \xi\right)+\frac{b_1^2}{2a_1}\lambda^2 +\zeta_{\lambda} \pi+\zeta_{\xi}\varpi.
\end{align}
Requiring the time preservation of the primary constraints $\pi \approx 0$ and $\varpi \approx 0$,
we obtain secondary constraints
\begin{align}
b_0+\frac{b_1}{a_1}p+b_2q+b_3\xi-\frac{b_1^2}{a_1}\lambda \approx 0, \quad a_3\xi-b_3\lambda \approx 0.
\end{align}
Requiring the time preservation of the secondary constraints, we obtain 
\begin{align} 
b_3\zeta_\xi - \frac{b_1^2}{a_1}\zeta_\lambda \approx \cdots , \quad
a_3\zeta_\xi-b_3\zeta_\lambda \approx 0. \label{ccsec}
\end{align}
Hence, if 
\be b_3+\f{a_3b_1^2}{a_1} \neq 0, \label{noter} \ee 
is satisfied, the consistency conditions~\eqref{ccsec} fix the Lagrange multipliers $\zeta_{\lambda},\zeta_{\xi}$ and thus no further constraint is obtained.
In this case we end up with four constraints and hence the system has 1 DOF.

Considering several special cases of the toy model \eqref{Ltoy} serve as pedagogical examples of application of the general argument in \S\ref{sec_aux21}, \ref{sec_aux22}, and \ref{sec_aux23}.
First, let us consider the case $a_3=b_3=0$, i.e.\
\begin{align} \label{Ltoy1}
    L=\frac{a_1}{2}\dot{q}^2-\frac{a_2}{2}q^2+\lambda(b_0+b_1 \dot{q} + b_2 q).
\end{align}
In this case the Lagrangian no longer depends on $\xi$.\footnote{The same Lagrangian is obtained even when $a_3\neq 0$ and $b_3=0$; the Euler-Lagrange equation of $\xi$ fixes $\xi=0$ and then \eqref{Ltoy1} is obtained after substituting it.} 
The variable of the system is $q$ only in addition to the Lagrange multiplier $\lambda$.
This case falls into the class we investigated in \S\ref{sec_aux22}.
As we clarified in \S\ref{sec_aux22}, the Ostrogradsky-like ghost exists when the conditions \ref{con1}, \ref{con2}, and \ref{con3} are satisfied. 
In the present case, the condition \ref{con1}: $a_1\ne 0$ is satisfied by the assumption.  
Furthermore, let us assume that the condition \ref{con2}: $b_1\neq 0$ is satisfied for a while.
Imposing $b_1 \neq 0$, the system does not have tertiary constraint as \eqref{noter} is satisfied. 
Thus, so far we assumed $a_3=b_3=0$, $a_1\ne 0$ and $b_1\neq 0$. 
The on-shell total Hamiltonian is then given by
\begin{align}
H_{\text{on-shell}}=\frac{a_2}{2}q^2-\frac{a_1}{2b_1^2}(b_0+b_2 q)^2-\frac{b_0+b_2q}{b_1}p
\,, \label{H_uns}
\end{align}
which is unbounded due to the last term linear in $p$, if $b_0\ne 0$ or $b_2\ne 0$.
Precisely, the condition \ref{con3} consists of these two cases, namely, \ref{con3}-a: $b_0\ne 0$, $b_1\ne 0$ and $b_2=0$, or \ref{con3}-b: $b_2\ne 0$.
Both cases \ref{con3}-a and \ref{con3}-b share the unbounded Hamiltonian due to the linear dependency of $p$, but by different reasons, as is explored in \S\ref{sec_aux22}.
If the condition \ref{con3}-a is satisfied, the constraint reads $\dot q=-b_0/b_1$, which does not allow stationary solution $\dot{q}=0$. 
Indeed, the on-shell total Hamiltonian $H_{\text{on-shell}}=\frac{a_2}{2}q^2-\frac{b_0}{b_1}p-\frac{a_1b_0^2}{2b_1^2}$ does not have stationary points. 
On the other hand, under the condition \ref{con3}-b, the on-shell total Hamiltonian has stationary points but all of them are saddle points. 
We can check the Hessian of the Hamiltonian at the stationary points
\begin{align}
    \bm{H}=\begin{pmatrix}
       0 & -\frac{b_2}{b_1}\\
       -\frac{b_2}{b_1} & a_2-\frac{a_1b_2^2}{b_1^2}
    \end{pmatrix}
    ,
\end{align}
which indeed has both negative and positive eigenvalues, if the condition \ref{con3}-b: $b_2\ne 0$ is satisfied.
In both cases \ref{con3}-a and \ref{con3}-b, we stress that here we obtained the unbounded Hamiltonian due to the linear momentum term despite the fact that the Lagrangian~\eqref{Ltoy1} is not equivalent to a higher-derivative theory.
While the Ostrogradsky theorem does not apply to this system, the conditions \ref{con1}--\ref{con3} serve as a more powerful tool to detect the pathology of the Hamiltonian.

The violation of at least one of the conditions \ref{con1}, \ref{con2}, or \ref{con3} is a necessary condition to evade the Ostrogradsky-like instability. First, when the condition \ref{con2} does not hold, i.e.\ $b_1=0$, the constraint is no longer nonholonomic and then the Ostrogradsky-like ghost does not exist. 
This case falls into the case discussed in \S\ref{sec_aux21}.
We also note that for $b_1=0$ the system possesses a tertiary constraint.

Second, the violation of the condition \ref{con3} means $b_0=b_2=0$. We then obtain a bounded Hamiltonian.
Note that the case $b_0=b_2=0$ (with $b_1,a_2 \neq 0$ and $a_3=b_3=0$) is nothing but the first-order formalism of the Lagrangian $L=\frac{1}{2a_2}\dot{\lambda}^2$ as we shall see below. 
In this case the Lagrangian~\eqref{Ltoy} reads
\begin{align} \label{Ltoy2}
    L=\frac{a_1}{2}\dot{q}^2-\frac{a_2}{2}q^2+b_1 \lambda \dot{q}.
\end{align}
By redefining the Lagrange multiplier
\begin{align}
    \lambda \rightarrow \frac{1}{b_1}\lambda-\frac{a_1}{2b_1}\dot{q},
\end{align}
we can absorb the kinetic term and obtain
\begin{align}
    L=\lambda\dot{q}-\frac{a_2}{2}q^2. \label{first_order}
\end{align}
The Euler-Lagrange equations for this Lagrangian are two first-order equations:
\begin{align}
    \dot{q}&=0, \label{foe1} \\
    \dot{\lambda}+a_2 q&=0. \label{foe2}
\end{align}
On the other hand, by taking integration by parts, the Lagrangian \eqref{first_order} becomes
\begin{align}
    L=-\dot{\lambda}q-\frac{a_2}{2}q^2,
\end{align}
for which we can eliminate $q$ by the use of its equation of motion. We then obtain
\begin{align}
    L=\frac{1}{2a_2}\dot{\lambda}^2.
\end{align}
of which the Euler-Lagrange equation is one second-order equation:
\begin{align}
    \ddot{\lambda}=0, \label{soe}
\end{align}
which is consistent with the system of the two first-order equations above. 
Indeed, taking a time derivative of \eqref{foe2} and using \eqref{foe1}, one can recover \eqref{soe}.

Third, we consider the case $a_3\neq 0$, which falls into the case we studied in \S\ref{sec_aux23}.
In this case,
the Lagrangian is degenerate in terms of the variables $q$ and $\xi$; that is, the violation of the condition \ref{con1}. 
Following the prescription given in \S\ref{sec_aux23}, we need to additionally require $b_3\neq 0$ to remove the Ostrogradsky-like instability since the nonholonomic constraint has to be solved in terms of the nondynamical variable. 
Indeed, 
in this case the general argument below \eqref{ccsec} applies, and 
the on-shell Hamiltonian is then given by
\begin{align}
    H_{\text{on-shell}}=\frac{1}{2a_1}p^2+\frac{a_2}{2}q^2+\frac{1}{2}\left(\frac{b_3^2}{a_3}-\frac{b_1^2}{a_1}\right)^{-1}\left(b_0+\frac{b_1}{a_1}p+b_2q\right)^2, \label{H_sta}
\end{align}
which can be bounded from below as far as $b_3\neq 0$. It should be contrasted with the unbounded Hamiltonian \eqref{H_uns} which we obtained under the assumption $b_3=0$.

It would be worthwhile stressing that the Hamiltonian \eqref{H_sta} can be bounded from below even if $a_1<0$. Let us consider a ghost Lagrangian with a wrong sign of the kinetic term
\begin{align}
    L_0=-\frac{1}{2}\dot{q}^2-\frac{1}{2}q^2
\end{align}
and try to remove the ghost instability by adding a constraint via the Lagrange multiplier without reducing the degree of freedom of $q$. One way is imposing the constraint $\dot{q}=0$, but this constraint only admits a constant $q$. Another way is introducing a nondynamical variable $\xi$ and imposing a constraint such as $b_1\dot{q}+b_2 q+b_3 \xi=0$ as discussed above. Thus, the ghost can be cured by adding a constraint to the system; however, it also requires a nondynamical variable in the second case. As discussed in \S\ref{sec_over1}, the second way is similar to construct a degenerate higher-order theory.

\subsection{Higher-derivative theories}%%%%%%%%%%%%%%%%%%%%%
\label{sec_ex2}

Let us study a higher-derivative Lagrangian with $N+1$-th order derivatives of a variable $\phi_0(t)$,
\begin{align}
    L=L(\phi_0^{(N+1)},\phi_0^{(N)},\cdots,\dot{\phi}_0,\phi_0),
\end{align}
of which an equivalent Lagrangian is
\begin{align}
    L_{\rm eq}=L(\dot{\phi}_{a},\dot{\phi}_0,\phi_0)+\lambda^a C_a,
    \quad a=1,2,\cdots N,
    \label{HD_eq}
\end{align}
with auxiliary variables $\phi_a$, Lagrange multipliers $\lambda^a$, and nonholonomic constraints 
\begin{align} \label{HD_const}
    C_a=\phi_a-\dot{\phi}_{a-1}.
\end{align}
In the equivalent form~\eqref{HD_eq}, the constraints are linear in the first-order derivative of all variables $\phi_{a},\phi_0$ except the highest one $\phi_{N}$. Therefore, for any Lagrangian, the Hessian
\begin{align}
    L_{IJ}=\frac{\partial^2 L}{\partial \dot \phi_I \partial \dot \phi_J}, \quad I,J=0,1,2,\cdots N,
\end{align}
can be nondegenerate by redefining the Lagrange multipliers along the same line as the toy model \eqref{L1_lam} if
\begin{align}
 L_{NN}\neq 0. \label{non-deg_cond}
 \end{align}
For instance, as for the Lagrangian
 \begin{align}
     L=\frac{1}{2}\ddot{\phi}_0^2,
\end{align}
the equivalent Lagrangian is
\begin{align}
    L_{\rm eq}=\frac{1}{2}\dot{\phi}_1^2+\lambda^1(\phi_1-\dot{\phi}_0),
\end{align}
which is a degenerate Lagrangian in terms of $\phi_1,\phi_0$.
However, we can transform it to a nondegenerate Lagrangian
\begin{align}
    L_{\rm eq}=\frac{1}{2}\dot{\phi}_1^2+\frac{1}{2}\dot{\phi}_0^2-\frac{1}{2}\phi_1^2 +\lambda^1(\phi_1-\dot{\phi}_0),
\end{align}
by redefining the Lagrange multiplier as
\begin{align}
    \lambda^1\rightarrow \lambda^1-\frac{1}{2}(\phi_1+\dot{\phi}_0).
\end{align}

The nondegeneracy condition~\eqref{non-deg_cond} corresponds to the criterion \ref{con1} for the existence of the Ostrogradsky-like instability found in \S\ref{sec_aux22}.
We can see that when \eqref{non-deg_cond} is satisfied, other criteria \ref{con2} and \ref{con3} are automatically satisfied. 
The Lagrange multipliers can be redefined such that the Hessian is given by
\begin{align}
    L_{IJ}={\rm diag}[1,1,\cdots,1,L_{NN}].
\end{align}
From \eqref{defMab} and \eqref{defCq} we then have 
\begin{align}
    M_{ab}=\delta_{ab}, \quad \frac{\partial C_a}{\partial \phi_I}=\delta_{aI},
\end{align}
that is,
\begin{align}
    {\rm det}(M_{ab})\neq 0, \quad \bm{C_q}\neq 0.
\end{align}
Thus, our criteria by means of the language of the nonholonomic constraints reduce to the Ostrogradsky's theorem: a higher-derivative Lagrangian which is  nondegenerate with respect to the highest-order derivatives leads to an unbounded Hamiltonian.

To conclude the existence of the Ostrogradsky ghost, only the nondegeneracy with respect to the highest derivative is important. However, the degeneracy with respect to the highest derivative term is the necessary condition but not a sufficient condition to evade the Ostrogradsky ghost~\cite{Motohashi:2014opa}. In the equivalent Lagrangian \eqref{HD_eq}, the degeneracy with respect to the highest derivative concludes $\phi_N$ is a nondynamical variable. Therefore, the constraint
\begin{align}
    \phi_N-\dot{\phi}_{N-1}=0
\end{align}
can be solved to fix $\phi_N=\dot{\phi}_{N-1}$ without introducing a higher derivative term of $\phi_{N-1}$. However, we still have $N-1$ nonholonomic constraints and then can conclude that the Hamiltonian is still unbounded. To remove all nonholonomic constraints, we require $N$ degeneracy conditions.

When there is only one variable $\phi_0$ in the original Lagrangian, it is almost trivial that the degeneracy of the highest derivative term is not a sufficient condition. For example, let us consider a Lagrangian with at most third derivative:
\begin{align}
    L=\frac{c_3}{2}\dddot{\phi}_0^2+\frac{c_2}{2}\ddot{\phi}_0^2+\frac{c_1}{2}\dot{\phi}_0^2,
\end{align}
which is a special case of the one considered in \cite{Motohashi:2017eya}.
The degeneracy of the highest derivative term means $c_3=0$. Even so, there is the second derivative term in the Lagrangian which leads to the Ostrogradsky ghost. We thus need to impose $c_2=0$ to remove the Ostrogradsky ghost. By using the Lagrange multipliers, the equivalent Lagrangian is
\begin{align}
    L_{\rm eq}=\frac{c_3}{2}\dot{\phi}_2^2+\frac{c_2}{2}\dot{\phi}_1^2+\frac{c_1}{2}\dot{\phi}_0^2
    +\lambda^1(\phi_1-\dot{\phi}_0)+\lambda^2(\phi_2-\dot{\phi}_1).
\end{align}
The condition $c_3=0$ leads to that $\phi_2$ is a nondynamical variable. After removing $\phi_2$ via solving the constraint $\phi_2-\dot{\phi}_1=0$, the condition $c_2=0$ corresponds to the condition that $\phi_1$ is nondynamical. Note that $\phi_1$ is never nondynamical unless the constraint $\phi_2-\dot{\phi}_1=0$ is solved. Therefore, the degeneracy conditions must be imposed sequentially: We have to first impose the degeneracy condition of the highest derivative part. After solving the nonholonomic constraint associated with the highest derivative in terms of the nondynamical variable, we then impose the degeneracy condition of the next highest derivative term. This procedure must be continued until all nonholonomic constraints are solved in terms of nondynamical variables.

Therefore, a concrete procedure to derive all degeneracy conditions is straightforward but complicated in multi-variable higher-derivative system.
This procedure was established in a series of works~\cite{Motohashi:2014opa,Motohashi:2016ftl,Motohashi:2017eya,Motohashi:2018pxg}. It has been shown that Ostrogradsky ghost of a Lagrangian with $N+1$-th order derivative of $\phi_0$ and with at most first order derivatives of $q^I$ can be removed by imposing $N$ degeneracy conditions. We thus do not discuss it in the present paper.

\subsection{Violation of the condition \ref{con1} or \ref{con2}}%%%%%%%%%%%%%%%%%%%%%
\label{sec_ex3}

Let us revisit the Ostrogradsky ghost-free system~\eqref{ex_intro5} 
\begin{align}
    L_{\rm Healthy}&=L_0+\lambda C,
    \nn 
    L_0&=\frac{\dot{Q}^2}{2}+\frac{\dot{\phi}^2}{2}-\frac{\xi^2}{4},
    \nn
    C&=Q-\dot{\phi}+\xi
    , \label{LH1}
\end{align}
As explained in \S\ref{sec_over1}, the degenerate higher-order theory is obtained by solving the constraint $Q-\dot{\phi}+\xi=0$ in terms of $Q$. In this subsection, on the other hand, we shall keep treating $(Q,\phi,\xi)$ as independent variables and discuss why the ghost is exorcised by means of the language of the (non)holonomic constraint.

The variables $Q,\phi$ are dynamical variables while $\xi$ is a solvable nondynamical variable and $\lambda$ is a Lagrange multiplier, respectively. However, one can regard the set $(\xi,\lambda)$ as solvable nondynamical variables since the matrix
\begin{align}
    \bm{D}=
    \begin{pmatrix}
       \frac{\partial^2 L}{\partial \xi^2} & \frac{\partial C}{\partial \xi} \\
        \frac{\partial C}{\partial \xi}  & 0
    \end{pmatrix}
    =
    \begin{pmatrix}
       -\frac{1}{2} & 1 \\
       1 & 0
    \end{pmatrix},
\end{align}
has the nonvanishing determinant, $\det\bm{D}=-1$, which satisfies the ghost-free criterion~\eqref{detD}. Hence, the nondynamical variables $(\xi,\lambda)$ are determined by the set of their Euler-Lagrange equations. Therefore, Theorem \ref{main_th} for systems with nonholonomic constraints cannot be applied to the system \eqref{LH1} and then it is free from the Ostrogradsky-like ghost.

In the first place, however, it is not clear which variables are nondynamical for a given $L_0$. The existence of a nondynamical variable is governed by the (non)degeneracy condition.
If we first interpret all variables $q^I=(Q,\phi,\xi)$ in $L_0$ as dynamical variables, the unconstrained Lagrangian $L_0$ is regarded as a degenerate Lagrangian. It can be thus understood as that the violation of the condition \ref{con1}, i.e.~the degeneracy, inhibits the appearance of the ghost in the Lagrangian \eqref{LH1}. This picture would be similar to the picture of the degenerate higher-order theory.

The system can be also seen as that the condition \ref{con2} is violated while the condition \ref{con1} holds when we change the variables. We define new Lagrange multiplier $\Lambda$ by
\begin{align}
    \lambda = \Lambda+\dot{\xi},
\end{align}
and obtain
\begin{align}
    L_{\rm Healthy}&=L_0'+\Lambda C,
    \nn
    L_0'&=\frac{\dot{Q}^2}{2}+\frac{\dot{\phi}^2}{2}-\frac{\xi^2}{4}+\dot{\xi}(Q-\dot{\phi}+\xi)
    . \label{LH2}
\end{align}
Due to the kinetic mixing $\dot{\xi}\dot{\phi}$ the unconstrained part $L_0'$ is no longer degenerate. Indeed, the kinetic matrix for $(Q,\phi,\xi)$ and its inverse are given by
\begin{align}
    L_{IJ}=
    \begin{pmatrix}
       1 & 0 & 0\\
       0 & 1 & -1 \\
       0 & -1 & 0  
    \end{pmatrix}
    ,\quad
    L^{IJ}=
    \begin{pmatrix}
       1 & 0 & 0 \\
       0 & 0 &-1 \\
       0& -1 & -1 
    \end{pmatrix}
    .
\end{align}
Thus, the condition \ref{con1} holds for \eqref{LH2}.
However, the condition \ref{con2} is now violated:
\begin{align}
    M\equiv C_I L^{IJ} C_J =\frac{\partial C}{\partial \dot{\phi}} L^{22} \frac{\partial C}{\partial \dot{\phi}}=0\,.
\end{align}
The violation of the condition \ref{con2} becomes manifest if we further introduce a variable $\Xi$ via
\begin{align}
    \xi=\Xi+\dot{\phi}.
\end{align}
The Lagrangian is then
\begin{align}
    L_{\rm Healthy}&=\frac{\dot{Q}^2}{2}+\frac{\dot{\phi}^2}{2}-\frac{(\Xi+\dot{\phi})^2}{4}-(\Xi+\dot{\phi})(\dot{Q}+\dot{\Xi})+\Lambda(Q+\Xi),
    \label{LH3}
\end{align}
where we have performed the integration by parts to eliminate the second derivative of $\phi$.  Now, the constraint $Q+\Xi\approx 0$ is just a holonomic constraint in terms of the variables $(Q,\phi,\Xi)$. Note that the kinetic matrix of \eqref{LH3} (or \eqref{LH2}) has positive and negative eigenvalues. Hence, there would exist a ghost degree of freedom if there were no constraint. However, this ghost degree of freedom is eliminated by the holonomic constraint $Q+\Xi=0$. Therefore, in the Lagrangian \eqref{LH3} (or \eqref{LH2}) it can be understood as that the ghost is exorcised as a result of a reduction of the phase space dimension by adding a constraint.

At least in this example, there is no essential difference between the violation of the condition~\ref{con1} and that of the condition~\ref{con2} since the difference is just in appearance. The point is that the Lagrangian $L_{\rm Healthy}=L_0(\dot{q}^I,q^I)+\lambda C(\dot{q}^I,q^I)$ with $q^I=(Q,\phi,\xi)$ only has two dynamical degree of freedom on shell. The ghost is evaded by the reduction of the phase space dimension.

\subsection{Violation of the condition \ref{con3}}%%%%%%%%%%%%%%%%%%%%%
\label{sec_ex4}

We shall consider two examples, without or with ghost, that violate the condition \ref{con3}.
The first toy model here is
\begin{align} \label{ex_vio3_GF}
    L=\frac{1}{2}(\dot{q}^1)^2+\frac{1}{2}(\dot{q}^2)^2-V(q^1,q^2)+\lambda[f_1(q^1,q^2)\dot{q}^1+f_2(q^1,q^2)\dot{q}^2],
\end{align}
for which the canonical momenta and the total Hamiltonian are
\begin{align}
    p_1&=\dot{q}^1+\lambda f_1,\quad p_2=\dot{q}^2+\lambda f_2, \quad 
    \pi=0,\\
    H_{\rm tot}&=\frac{1}{2}p_1^2-p_1 \lambda f_1+\frac{1}{2}\lambda^2 f_1^2+\frac{1}{2}p_2^2-p_2 \lambda f_2+\frac{1}{2}\lambda^2 f_2^2+V+\zeta_{\lambda} \pi.
\end{align}
As far as $f_1^2+f_2^2\neq 0$, there are only the primary and the secondary constraints,
\begin{align}
    \pi \approx 0,\quad \lambda \approx \frac{p_1f_1+p_2 f_2}{f_1^2+f_2^2},
\end{align}
and then the phase space dimension is four. Namely, the nonholonomic constraint does not reduce the phase space dimension of $(q^I,p_I)$. 
However, our theorem does not apply to this system because the condition \ref{con3} is violated: the stationary point $\dot{q}^1=\dot{q}^2=0$ is allowed since it is not in contradiction with the constraint, and also $\bm{C_q}$ vanishes at the stationary point. Indeed, the on-shell Hamiltonian is given by
\begin{align}
    H_\text{on-shell}=\frac{(p_1f_2-p_2f_1)^2}{2(f_1^2+f_2^2)}+V,
\end{align}
which is bounded from below as far as the potential is bounded. The on-shell Hamiltonian has a minimum on the hypersurface $p_1f_2-p_2 f_1=0$ which corresponds to $\dot{q}^1=\dot{q}^2=0$ under $f_1,f_2\neq 0$.

Next example is the Lagrangian,
\begin{align} \label{ex_vio3_ghost}
    L=\frac{1}{2}(\dot{q}^1)^2+\frac{1}{2}(\dot{q}^2)^2-V(q^1,q^2)+\lambda[\dot{q}^2-(q^1)^2]
\end{align}
where the potential $V$ is supposed to have a local minimum at $q^1=0$. 
The condition \ref{con3} is violated since there exists a stationary solution $\dot{q}^2=0$ at $q^1=0$, and also $\bm{C_q}$ vanishes there.  
Does the on-shell Hamiltonian has a local minimum there? The canonical momenta are
\begin{align}
    p_1=\dot{q}^1,\quad p_2=\dot{q}^2+\lambda,
\end{align}
and all constraints are
\begin{align}
    \pi\approx 0, \quad p_2-\lambda-(q^1)^2\approx 0.
\end{align}
The on-shell Hamiltonian is then
\begin{align}
    H_\text{on-shell}=\frac{1}{2}p_1^2+V+p_2(q^1)^2-\frac{1}{2}(q^1)^4.
\end{align}
Hence, the point $q^1=p_2=0$ ($\Leftrightarrow \dot{q}^2=q^1=0)$ is not a local maximum but just a saddle point. The Hamiltonian is unbounded from below due the linear dependency of $p_2$ (the last term is not essential for the boundedness of the Hamiltonian when the potential increases faster than $\frac{1}{2}(q^1)^4$).

Therefore, in general, if the condition \ref{con3} is violated, there is a possibility to have a healthy Hamiltonian; however, an additional analysis is required to conclude whether the stationary point is indeed a local minimum or just a saddle point. The stationary point is a local minimum if the Hessian is positive definite but the stationary point can be either a local minimum or a saddle point if the Hessian is positive semi-definite. The additional analysis is beyond the scope of the present paper.

\subsection{On a bounded Hamiltonian without local minimum}
\label{sec_canonical}
As we stressed, strictly speaking,
our theorem clarifies the absence of local minimum of the Hamiltonian except the boundary of the domain of the phase space $(q^I,p_I)$, but it does not necessarily mean that the Hamiltonian is unbounded since the Hamiltonian is still allowed to take a finite value at the boundary. This happens if the Hamiltonian takes a form like the hyperbolic tangent 
or if the domain of the variables is compact. Indeed, the authors of \cite{Ganz:2018mqi} (see also \cite{Chaichian:2014qba}) argued that in the context of mimetic gravity a positive definite energy may be realized by restricting the field domain even if the Hamiltonian is linear in a canonical variable. Let us study concrete examples of such systems here.

The simplest example of a bounded Hamiltonian without local minimum would be a Hamiltonian of the harmonic oscillator after a canonical transformation (see e.g.\ \cite{goldstein2002classical} \S 9.3 The Harmonic Oscillator, page 377). 
The standard form of the Hamiltonian of the harmonic oscillator is $H=\frac{1}{2}p^2+\frac{\omega^2}{2}q^2$. By means of a generating function $W=-\frac{1}{2\omega}p^2\tan Q$
the Hamiltonian is transformed into $H=\omega P$ where $(q,p)$ and $(Q,P)$ are related by
\begin{align}
    q=-\frac{\partial W}{\partial p}=\frac{p}{\omega} \tan Q\,, \quad
    P=-\frac{\partial W}{\partial Q}=\frac{p^2}{2\omega} \frac{1}{\cos^2 Q}
    .
\end{align}
It is clear in this form that $Q$ is cyclic and hence the conjugate momentum $P$ remains a positive constant corresponding to the energy of the system. 
Although $H$ linearly depends on the canonical momentum $P$, the Hamiltonian is bounded from below because $P$ is positive definite. Note however that the canonical transformation is singular at $P=0$ and then the domain of $P$ is $P>0$ in which $H$ has no stationary point.
One should use $(q,p)$ instead of $(Q,P)$ to analyze the physical properties of the point $P=0$ properly.  Needless to say, in this example, the point $P=0$ (zero energy) is not physically allowed at the quantum level due to the zero point energy of the harmonic oscillator and thus the domain $P>0$ is sufficient. Nonetheless, without the knowledge of the equivalence between the Hamiltonian $H=\omega P$ with $P>0$ and that of the harmonic oscillator, one may wonder why the domain $P\leq 0$ is removed and what happens at the boundary $P=0$.

In our analysis, we have concluded that the Hamiltonian of a nonholonomic system has no local minimum except the boundary of the domain. If the Hamiltonian is finite at the boundary and the system is well-defined there, the boundary should not be a physical singularity and then there would be a canonical transformation to cover the boundary after which the Hamiltonian may have a minimum inside the domain (the transformation from $(Q,P)$ to $(q,p)$ in the above example). If this is the case, after the Legendre transformation, the resultant Lagrangian must not have a nonholonomic constraint; that is, the nonholonomic constraint is converted into a holonomic one via a canonical transformation. If this is possible, this would be a generalization of a way out from a pathological Hamiltonian discussed in \S\ref{sec_aux23}. 

As an example we reconsider the Lagrangian
\be L= \frac{1}{2}(\dot{Q}^2-\omega^2) + \Lambda (\dot{Q}-\omega), 
\label{with_kinetic}
\ee
where $\omega$ is a positive constant.
This Lagrangian is a class of \eqref{Ltoy} (except the constant term $-\omega^2/2$) where we use a different notation than \eqref{Ltoy} in order to directly see the analogy to the harmonic oscillator. As shown, the Hamiltonian of this system does not admit a local minimum. For our purpose, it is useful to redefine the Lagrange multiplier $\Lambda \rightarrow \Lambda-(\dot{Q}+\omega)/2$ in order to write the Lagrangian as
\begin{align}
    L=\Lambda (\dot{Q}-\omega)
    . \label{eq_harmonic}
\end{align}
In the form \eqref{eq_harmonic}, there are a couple of primary constraints, and the corresponding Hamiltonian is
\begin{align}
    H_{\rm tot}=\omega P +\zeta^1 \Pi + \zeta^2 (P-\Lambda)
    , \label{H_harmonic}
\end{align}
where $\zeta^1,\zeta^2$ are Lagrange multipliers and $P, \Pi$ are the canonical momenta of $Q,\Lambda$, respectively. The on-shell Hamiltonian has no local minimum in the domain $-\infty < P < \infty$. However, the on-shell Hamiltonian takes the same form as the harmonic oscillator suggesting no pathology when $P>0$.\footnote{The consistency conditions of the constraints fix the Lagrange multipliers as $\zeta^1=\zeta^2=0$, which can be substituted into the total Hamiltonian. The Hamiltonian of the harmonic oscillator is then obtained. However, we shall retain $\zeta^1$ and $\zeta^2$ to see the conversion of a nonholonomic constraint into a holonomic one.} Hence, let us suppose $P>0$ which implies $\Lambda>0$ in terms of the configuration space variables due to the constraint $P-\Lambda \approx 0$. To analyze the point $P=0$, we take the canonical transformation $(Q,\Lambda,P,\Pi) \rightarrow (q,\lambda,p,\pi)$ via the generating function 
\begin{align}
    W'=\frac{1}{2\omega}p^2 \tan Q + \Lambda \pi
    ,
\end{align}
which yield 
\begin{align}
    q=\frac{\partial W'}{\partial p}=\frac{p}{\omega} \tan Q\,, \quad
    P=\frac{\partial W'}{\partial Q}=\frac{p^2}{2\omega} \frac{1}{\cos^2 Q}\,, \quad
    \lambda=\frac{\partial W'}{\partial \pi}=\Lambda\,, \quad
    \Pi=\frac{\partial W'}{\partial \Lambda}=\pi
    .
\end{align}
After the canonical transformation, the total Hamiltonian is given by
\be H_{\rm tot}=\f{1}{2}p^2 + \f{\omega^2}{2}q^2 +\zeta^1 \pi + \zeta^2 \mk{\f{1}{2\omega}p^2 + \f{\omega}{2}q^2-\lambda},\ee
We can take the Legendre transformation to obtain the corresponding Lagrangian. By using the Hamilton equation
\begin{align}
    \dot{q}=p\left(  1+\frac{\zeta_2}{\omega} \right),
    \quad
    % \dot{\pi}
    \dot\lambda=\zeta^1
    ,
\end{align}
we obtain
\begin{align}
    L=\dot{q}p+\dot{\lambda}\pi-H_{\rm tot}=\frac{\omega}{2(\zeta^2+\omega)}\dot{q}^2-\frac{1}{2}\omega(\zeta^2+\omega)q^2+\lambda \zeta^2
    .
\end{align}
Clearly, the constraint equation implemented by the Lagrange multiplier $\lambda$ is holonomic. Substituting the solution $\zeta^2=0$ into the Lagrangian, we finally obtain the Lagrangian of the harnomic oscillator
\begin{align}
    L=\frac{1}{2}\dot{q}^2-\frac{\omega^2}{2} q^2
    . \label{harmonic}
\end{align}
This implies that the system \eqref{eq_harmonic} with $\Lambda>0$ is equivalent to the harmonic oscillator with a non-zero energy and the ``boundary'' $\Lambda=0$ is naturally interpreted as the zero energy state of the harmonic oscillator. Again, at the quantum level, the physical states require $P>0~(\Lambda>0)$ due to the zero-point energy. Therefore, the ad hoc assumption $P>0$ may be justified by the equivalence to the harmonic oscillator.

Therefore, while the Lagrangian \eqref{eq_harmonic} with $-\infty< \Lambda < \infty$ is pathological, the same Lagrangian with the restricted domain $0<\Lambda<\infty$ has no pathology. At the quantum level the singularity of the canonical transformation, $\Lambda=0$, is not physically allowed which would guarantee the equivalence between \eqref{eq_harmonic} with $\Lambda>0$ and \eqref{harmonic}. Although this example is so simple, we may expect that the same trick can be applied to another unbounded Hamiltonian in order to obtain a healthy system (see e.g.~the argument of \cite{Ganz:2018mqi}). We leave the general analysis for a future study.

\section{Conclusion}%%%%%%%%%%%%%%%%%%%%%%%%%%%%%%%%%%%%%%%%%
\label{sec_conc}

Constraints have been playing a central role to exorcise the Ostrogradsky ghosts associated with higher derivatives.  In this paper, however, we have clarified that adding constraints to a system is not always a good thing, and can summon ghost degrees of freedom as highlighted in \S\ref{sec_over} by a simple example.
Such ghosts may or may not be associated with higher derivatives, and hence our result includes the Ostrogradsky theorem as a special case.  

In general, constraints show up when a Lagrangian contains nondynamical variables, whose derivatives do not appear in Lagrangian.  
Nondynamical variables are qualitatively different, depending on whether they appear in Lagrangian nonlinearly or linearly.  For the former case, one can in principle write down the nondynamical variables in terms of other variables by solving their Euler-Lagrange equations.  On the other hand, the latter case corresponds to the constraints implemented by Lagrange multipliers, which we found has a rich structure from the point of view of un/boundedness of the Hamiltonian.
  
We focused on theories with solvable auxiliary variables in \S\ref{sec_aux1}, and derived degeneracy condition~\eqref{deg} to evade the Ostrogradsky ghost.  The advantage of the degeneracy condition~\eqref{deg} is that one can check it without substituting solutions of auxiliary variables explicitly.  It allows us a wide range of models for application since in general the Euler-Lagrange equations for the auxiliary variables may be solved only locally.
  
In \S\ref{sec_aux2}, we have investigated systems having constraints with Lagrange multipliers.  
This case further divided into two cases depending on whether the constraints are holonomic (velocity-independent) or nonholonomic (velocity-dependent).
Linearly independent set of holonomic constraints on dynamical variables always reduces the phase space dimension of dynamical variables, whereas in general nonholonomic constraints do not. 
We have clarified that adding nonholonomic constraints that does not reduce the phase space dimension in general leads to a pathological Hamiltonian due to the indefiniteness of the Hessian.  This occurs even if the original Lagrangian before adding the constraints is healthy, and/or if the resultant Lagrangian is not equivalent to higher-derivative one.

More precisely, we have established a set of sufficient conditions for 
the absence of local extremum of
Hamiltonian as the conditions~\ref{con1}--\ref{con3} given in \S\ref{sec_aux22}, as a generalization of the Ostrogradsky theorem.  Their physical meaning is reasonable as the condition~\ref{con1}: nondegeneracy, the condition~\ref{con2}: nonholonomic constraints not reducing the phase space dimension, the condition~\ref{con3}-a: no stationary points, and the condition~\ref{con3}-b: all stationary points are saddle points.  
For higher-derivative theories, only the condition~\ref{con1} is relevant, which is precisely related to the nondegeneracy assumption of the Ostrogradsky theorem, and the conditions~\ref{con2} and \ref{con3}-b are automatically satisfied.  However, there are various ways to satisfy these conditions even if the model is not associated with higher-derivative theories.  We have considered such examples in \S\ref{sec_aux22}, \S\ref{sec_aux24} and \S\ref{sec_ex1}.  Thus, our theorem is a natural generalization of the Ostrogradsky theorem.

The violation of either of the condition~\ref{con1}--\ref{con3} is necessary to evade the absence of local extremum of
Hamiltonian but not always sufficient.  In \S\ref{sec_aux23}, \S\ref{sec_aux24}, \S\ref{sec_ex}, we provided various examples to highlight the application and limitation of our theorem.  

In particular, in \S\ref{sec_aux23} we have clarified that a possible way out from the pathological Hamiltonian is to convert nonholonomic constraints to holonomic ones, or more generally, Lagrange multipliers to solvable auxiliary variables, by introducing auxiliary variables. Such a process is possible if the Lagrangian has at least the same number of the auxiliary variables as the number of the nonholonomic constraints and satisfies the ghost-free criterion~\eqref{detD}. We can then solve the constraints for the auxiliary variables and erase them by substituting the solutions, following the prescription in \S\ref{sec_aux21}.
This process embeds the interplay between degenerate higher-order theory and lower-order theory with auxiliary variables related through an invertible transformation, 
which has been focused in the context of construction of higher-derivative theories of modified gravity, 
into a broader context.

In \S\ref{sec_canonical} we also argue another way out from the pathological Hamiltonian by the following trick. Let us consider a Hamiltonian without local minimum, say \eqref{H_harmonic}. We first restrict the domain of the canonical variables to bound the Hamiltonian, and then take a canonical transformation to guarantee the positive definiteness. The Hamiltonian admits a local minimum at the point that was originally the boundary of the restricted domain. Since the transformation is singular at the boundary of the restricted domain, one may worry about the equivalence between two systems. However, as for the example \eqref{H_harmonic}, the transformed theory is the harmonic oscillator and thus the singular point is not physically allowed by the zero-point energy at the quantum level. Hence, two systems are equivalent in the physical domain and the ad hoc assumption, the restriction of the domain, is justified.
This argument on the bounded Hamiltonian with a restricted domain should be related to the instability issue of the mimetic gravity in which theory the kinetic term of a scalar field is imposed to be constant by the mimetic constraint. The constraint of \eqref{with_kinetic}, $\dot{Q}-\omega =0$, could be regarded as a toy model of the mimetic constraint. It must be interesting to explore whether the same trick can be justified in the mimetic gravity and in more generic theories with unbounded Hamiltonian (see \cite{Ganz:2018mqi} for the discussion on the mimetic gravity where they study the replacement $\Lambda \rightarrow e^{\Lambda}$ instead of the canonical transformation).

In summary, our approach provides a unified way to detect and eliminate the Ostrogradsky ghost.
In the previous works, the origin of the pathology was regarded to be associated with higher derivatives. 
However, we have clarified that it is more generally attributed to nonholonomic constraints that do not reduce the phase space dimension. 
Such ghosts can be evaded by introducing auxiliary variables in a proper way, or restricting the domain of the canonical variables.

A subtle point is the freedom of redefinition of Lagrange multipliers. As discussed in \S\ref{sec_aux24} and \S\ref{sec_ex3}, the conditions are not invariant under an invertible transformation of the variables $(q^I,\xi^i,\lambda^a)$. It may be possible to improve our conditions in a covariant manner.
Also, throughout the present paper, we restricted ourselves to analytical mechanics of interacting point particles.  
It would be intriguing to generalize our arguments to field theory.
In particular, auxiliary variables play an essential role non only in theories of mimetic gravity and but also in the metric-affine (Palatini) formalism of higher derivative scalar-tensor theories~\cite{Aoki:2018lwx,Aoki:2019rvi,Helpin:2019kcq}. Generalization of our result to theories of gravity must be helpful to extract the essential properties of auxiliary variables in such theories and then to extend and/or constrain theories of modified gravity.
We leave these issues for a future work.

\acknowledgments%%%%%%%%%%%%%%%%%%%%%%%%%%%%%%%%%%%%%%%% 
This work was supported in part by Japan Society for the Promotion of Science (JSPS) Grants-in-Aid for Scientific Research (KAKENHI)  No.\ JP19J00895 (K.A.), No.\ JP17H06359 (H.M.), No.\ JP18K13565 (H.M.).

\bibliography{ref}

\providecommand{\href}[2]{#2}\begingroup\raggedright\begin{thebibliography}{10}

\bibitem{Clifton:2011jh}
T.~Clifton, P.~G. Ferreira, A.~Padilla and C.~Skordis, \emph{{Modified Gravity
  and Cosmology}},
  \href{https://doi.org/10.1016/j.physrep.2012.01.001}{\emph{Phys. Rept.}
  {\bfseries 513} (2012) 1} [\href{https://arxiv.org/abs/1106.2476}{{\ttfamily
  1106.2476}}].

\bibitem{Joyce:2014kja}
A.~Joyce, B.~Jain, J.~Khoury and M.~Trodden, \emph{{Beyond the Cosmological
  Standard Model}},
  \href{https://doi.org/10.1016/j.physrep.2014.12.002}{\emph{Phys. Rept.}
  {\bfseries 568} (2015) 1} [\href{https://arxiv.org/abs/1407.0059}{{\ttfamily
  1407.0059}}].

\bibitem{Berti:2015itd}
E.~Berti et~al., \emph{{Testing General Relativity with Present and Future
  Astrophysical Observations}},
  \href{https://doi.org/10.1088/0264-9381/32/24/243001}{\emph{Class. Quant.
  Grav.} {\bfseries 32} (2015) 243001}
  [\href{https://arxiv.org/abs/1501.07274}{{\ttfamily 1501.07274}}].

\bibitem{Koyama:2015vza}
K.~Koyama, \emph{{Cosmological Tests of Modified Gravity}},
  \href{https://doi.org/10.1088/0034-4885/79/4/046902}{\emph{Rept. Prog. Phys.}
  {\bfseries 79} (2016) 046902}
  [\href{https://arxiv.org/abs/1504.04623}{{\ttfamily 1504.04623}}].

\bibitem{Ostrogradsky:1850fid}
M.~Ostrogradsky, \emph{{Memoires sur les equations differentielles, relatives
  au probleme des isoperimetres}}, {\emph{Mem. Acad. St. Petersbourg}
  {\bfseries 6} (1850) 385}.

\bibitem{Woodard:2015zca}
R.~P. Woodard, \emph{{Ostrogradsky's theorem on Hamiltonian instability}},
  \href{https://doi.org/10.4249/scholarpedia.32243}{\emph{Scholarpedia}
  {\bfseries 10} (2015) 32243}
  [\href{https://arxiv.org/abs/1506.02210}{{\ttfamily 1506.02210}}].

\bibitem{Raidal:2016wop}
M.~Raidal and H.~Veermäe, \emph{{On the Quantisation of Complex Higher
  Derivative Theories and Avoiding the Ostrogradsky Ghost}},
  \href{https://doi.org/10.1016/j.nuclphysb.2017.01.024}{\emph{Nucl.\ Phys.\ B}
  {\bfseries 916} (2017) 607}
  [\href{https://arxiv.org/abs/1611.03498}{{\ttfamily 1611.03498}}].

\bibitem{Smilga:2017arl}
A.~Smilga, \emph{{Classical and quantum dynamics of higher-derivative
  systems}}, \href{https://doi.org/10.1142/S0217751X17300253}{\emph{Int.\ J.\
  Mod.\ Phys.\ A} {\bfseries 32} (2017) 1730025}
  [\href{https://arxiv.org/abs/1710.11538}{{\ttfamily 1710.11538}}].

\bibitem{Motohashi:2020psc}
H.~Motohashi and T.~Suyama, \emph{{Quantum Ostrogradsky theorem}},
  \href{https://arxiv.org/abs/2001.02483}{{\ttfamily 2001.02483}}.

\bibitem{Motohashi:2014opa}
H.~Motohashi and T.~Suyama, \emph{{Third order equations of motion and the
  Ostrogradsky instability}},
  \href{https://doi.org/10.1103/PhysRevD.91.085009}{\emph{Phys. Rev.}
  {\bfseries D91} (2015) 085009}
  [\href{https://arxiv.org/abs/1411.3721}{{\ttfamily 1411.3721}}].

\bibitem{Chen:2012au}
T.-j. Chen, M.~Fasiello, E.~A. Lim and A.~J. Tolley, \emph{{Higher derivative
  theories with constraints: Exorcising Ostrogradski's Ghost}},
  \href{https://doi.org/10.1088/1475-7516/2013/02/042}{\emph{JCAP} {\bfseries
  1302} (2013) 042} [\href{https://arxiv.org/abs/1209.0583}{{\ttfamily
  1209.0583}}].

\bibitem{Langlois:2015cwa}
D.~Langlois and K.~Noui, \emph{{Degenerate higher derivative theories beyond
  Horndeski: evading the Ostrogradski instability}},
  \href{https://doi.org/10.1088/1475-7516/2016/02/034}{\emph{JCAP} {\bfseries
  1602} (2016) 034} [\href{https://arxiv.org/abs/1510.06930}{{\ttfamily
  1510.06930}}].

\bibitem{Motohashi:2016ftl}
H.~Motohashi, K.~Noui, T.~Suyama, M.~Yamaguchi and D.~Langlois, \emph{{Healthy
  degenerate theories with higher derivatives}},
  \href{https://doi.org/10.1088/1475-7516/2016/07/033}{\emph{JCAP} {\bfseries
  1607} (2016) 033} [\href{https://arxiv.org/abs/1603.09355}{{\ttfamily
  1603.09355}}].

\bibitem{Motohashi:2017eya}
H.~Motohashi, T.~Suyama and M.~Yamaguchi, \emph{{Ghost-free theory with
  third-order time derivatives}},
  \href{https://doi.org/10.7566/JPSJ.87.063401}{\emph{J. Phys. Soc. Jap.}
  {\bfseries 87} (2018) 063401}
  [\href{https://arxiv.org/abs/1711.08125}{{\ttfamily 1711.08125}}].

\bibitem{Motohashi:2018pxg}
H.~Motohashi, T.~Suyama and M.~Yamaguchi, \emph{{Ghost-free theories with
  arbitrary higher-order time derivatives}},
  \href{https://doi.org/10.1007/JHEP06(2018)133}{\emph{JHEP} {\bfseries 06}
  (2018) 133} [\href{https://arxiv.org/abs/1804.07990}{{\ttfamily
  1804.07990}}].

\bibitem{BenAchour:2016fzp}
J.~Ben~Achour, M.~Crisostomi, K.~Koyama, D.~Langlois, K.~Noui and G.~Tasinato,
  \emph{{Degenerate higher order scalar-tensor theories beyond Horndeski up to
  cubic order}}, \href{https://doi.org/10.1007/JHEP12(2016)100}{\emph{JHEP}
  {\bfseries 12} (2016) 100}
  [\href{https://arxiv.org/abs/1608.08135}{{\ttfamily 1608.08135}}].

\bibitem{Chamseddine:2013kea}
A.~H. Chamseddine and V.~Mukhanov, \emph{{Mimetic Dark Matter}},
  \href{https://doi.org/10.1007/JHEP11(2013)135}{\emph{JHEP} {\bfseries 11}
  (2013) 135} [\href{https://arxiv.org/abs/1308.5410}{{\ttfamily 1308.5410}}].

\bibitem{Lim:2010yk}
E.~A. Lim, I.~Sawicki and A.~Vikman, \emph{{Dust of Dark Energy}},
  \href{https://doi.org/10.1088/1475-7516/2010/05/012}{\emph{JCAP} {\bfseries
  1005} (2010) 012} [\href{https://arxiv.org/abs/1003.5751}{{\ttfamily
  1003.5751}}].

\bibitem{Gao:2010gj}
C.~Gao, Y.~Gong, X.~Wang and X.~Chen, \emph{{Cosmological models with Lagrange
  Multiplier Field}},
  \href{https://doi.org/10.1016/j.physletb.2011.06.085}{\emph{Phys. Lett.}
  {\bfseries B702} (2011) 107}
  [\href{https://arxiv.org/abs/1003.6056}{{\ttfamily 1003.6056}}].

\bibitem{Capozziello:2010uv}
S.~Capozziello, J.~Matsumoto, S.~Nojiri and S.~D. Odintsov, \emph{{Dark energy
  from modified gravity with Lagrange multipliers}},
  \href{https://doi.org/10.1016/j.physletb.2010.08.030}{\emph{Phys. Lett.}
  {\bfseries B693} (2010) 198}
  [\href{https://arxiv.org/abs/1004.3691}{{\ttfamily 1004.3691}}].

\bibitem{Sebastiani:2016ras}
L.~Sebastiani, S.~Vagnozzi and R.~Myrzakulov, \emph{{Mimetic gravity: a review
  of recent developments and applications to cosmology and astrophysics}},
  \href{https://doi.org/10.1155/2017/3156915}{\emph{Adv. High Energy Phys.}
  {\bfseries 2017} (2017) 3156915}
  [\href{https://arxiv.org/abs/1612.08661}{{\ttfamily 1612.08661}}].

\bibitem{Takahashi:2017pje}
K.~Takahashi and T.~Kobayashi, \emph{{Extended mimetic gravity: Hamiltonian
  analysis and gradient instabilities}},
  \href{https://doi.org/10.1088/1475-7516/2017/11/038}{\emph{JCAP} {\bfseries
  11} (2017) 038} [\href{https://arxiv.org/abs/1708.02951}{{\ttfamily
  1708.02951}}].

\bibitem{Langlois:2018jdg}
D.~Langlois, M.~Mancarella, K.~Noui and F.~Vernizzi, \emph{{Mimetic gravity as
  DHOST theories}},
  \href{https://doi.org/10.1088/1475-7516/2019/02/036}{\emph{JCAP} {\bfseries
  02} (2019) 036} [\href{https://arxiv.org/abs/1802.03394}{{\ttfamily
  1802.03394}}].

\bibitem{Ramazanov:2016xhp}
S.~Ramazanov, F.~Arroja, M.~Celoria, S.~Matarrese and L.~Pilo, \emph{{Living
  with ghosts in Ho\v rava-Lifshitz gravity}},
  \href{https://doi.org/10.1007/JHEP06(2016)020}{\emph{JHEP} {\bfseries 06}
  (2016) 020} [\href{https://arxiv.org/abs/1601.05405}{{\ttfamily
  1601.05405}}].

\bibitem{Ganz:2018mqi}
A.~Ganz, P.~Karmakar, S.~Matarrese and D.~Sorokin, \emph{{Hamiltonian analysis
  of mimetic scalar gravity revisited}},
  \href{https://doi.org/10.1103/PhysRevD.99.064009}{\emph{Phys. Rev. D}
  {\bfseries 99} (2019) 064009}
  [\href{https://arxiv.org/abs/1812.02667}{{\ttfamily 1812.02667}}].

\bibitem{Zumalacarregui:2013pma}
M.~Zumalacárregui and J.~García-Bellido, \emph{{Transforming gravity: from
  derivative couplings to matter to second-order scalar-tensor theories beyond
  the Horndeski Lagrangian}},
  \href{https://doi.org/10.1103/PhysRevD.89.064046}{\emph{Phys. Rev.}
  {\bfseries D89} (2014) 064046}
  [\href{https://arxiv.org/abs/1308.4685}{{\ttfamily 1308.4685}}].

\bibitem{Aoki:2018zcv}
K.~Aoki, C.~Lin and S.~Mukohyama, \emph{{Novel matter coupling in general
  relativity via canonical transformation}},
  \href{https://doi.org/10.1103/PhysRevD.98.044022}{\emph{Phys. Rev.}
  {\bfseries D98} (2018) 044022}
  [\href{https://arxiv.org/abs/1804.03902}{{\ttfamily 1804.03902}}].

\bibitem{Takahashi:2017zgr}
K.~Takahashi, H.~Motohashi, T.~Suyama and T.~Kobayashi, \emph{{General
  invertible transformation and physical degrees of freedom}},
  \href{https://doi.org/10.1103/PhysRevD.95.084053}{\emph{Phys. Rev.}
  {\bfseries D95} (2017) 084053}
  [\href{https://arxiv.org/abs/1702.01849}{{\ttfamily 1702.01849}}].

\bibitem{Gabadadze:2012tr}
G.~Gabadadze, K.~Hinterbichler, J.~Khoury, D.~Pirtskhalava and M.~Trodden,
  \emph{{A Covariant Master Theory for Novel Galilean Invariant Models and
  Massive Gravity}},
  \href{https://doi.org/10.1103/PhysRevD.86.124004}{\emph{Phys. Rev.}
  {\bfseries D86} (2012) 124004}
  [\href{https://arxiv.org/abs/1208.5773}{{\ttfamily 1208.5773}}].

\bibitem{Meyer:2010}
C.~D. Meyer, \emph{{Matrix Analysis and Applied Linear Algebra}}. Society for
  Industrial and Applied Mathematics, 2010.

\bibitem{Chaichian:2014qba}
M.~Chaichian, J.~Kluson, M.~Oksanen and A.~Tureanu, \emph{{Mimetic dark matter,
  ghost instability and a mimetic tensor-vector-scalar gravity}},
  \href{https://doi.org/10.1007/JHEP12(2014)102}{\emph{JHEP} {\bfseries 12}
  (2014) 102} [\href{https://arxiv.org/abs/1404.4008}{{\ttfamily 1404.4008}}].

\bibitem{goldstein2002classical}
H.~Goldstein, C.~Poole and J.~Safko, \emph{Classical Mechanics}. Addison
  Wesley, 2002.

\bibitem{Aoki:2018lwx}
K.~Aoki and K.~Shimada, \emph{{Galileon and generalized Galileon with
  projective invariance in a metric-affine formalism}},
  \href{https://doi.org/10.1103/PhysRevD.98.044038}{\emph{Phys. Rev.}
  {\bfseries D98} (2018) 044038}
  [\href{https://arxiv.org/abs/1806.02589}{{\ttfamily 1806.02589}}].

\bibitem{Aoki:2019rvi}
K.~Aoki and K.~Shimada, \emph{{Scalar-metric-affine theories: Can we get
  ghost-free theories from symmetry?}},
  \href{https://doi.org/10.1103/PhysRevD.100.044037}{\emph{Phys. Rev.}
  {\bfseries D100} (2019) 044037}
  [\href{https://arxiv.org/abs/1904.10175}{{\ttfamily 1904.10175}}].

\bibitem{Helpin:2019kcq}
T.~Helpin and M.~S. Volkov, \emph{{Varying the Horndeski Lagrangian within the
  Palatini approach}},
  \href{https://doi.org/10.1088/1475-7516/2020/01/044}{\emph{JCAP} {\bfseries
  2001} (2020) 044} [\href{https://arxiv.org/abs/1906.07607}{{\ttfamily
  1906.07607}}].

\end{thebibliography}\endgroup
\bibliographystyle{JHEP}

\end{document}